\documentclass[iop,apj]{emulateapj}
\pdfoutput=1
\usepackage{graphicx}
\usepackage{natbib}
\usepackage{amsmath}
\newcommand{\refeq}[1]{(\ref{#1})}
\newcommand{\stx}[1]{{*,\mathrm{#1}}}
\newlength{\floatwidth}
\begin{document}
\title{Shapes and Probabilities of Galaxy Clusters}
\author{Abel Yang}
\affil{Department of Astronomy, University of Virginia, Charlottesville, VA 22904}
\author{William C. Saslaw}
\affil{Institute of Astronomy, Madingley Road, Cambridge CB3 0HA, UK; and Department of Astronomy, University of Virginia, Charlottesville, VA 22904}
\begin{abstract}
We develop a general theory for estimating the probability that a galaxy cluster of a given shape exists. The theory is based on the observed result that the distribution of galaxies is very close to quasi-equilibrium, in both its linear and nonlinear regimes. This places constraints on the spatial configuration of a cluster of galaxies in quasi-equilibrium. In particular, we show that that a cluster of galaxies may be described as a collection of nearly virialized subclusters of approximately the same mass. Clusters that contain more than 10 subclusters are very likely to be completely virialized. Using our theory, we develop a method for comparing probabilities of different spatial configurations of subclusters. As an illustrative example, we show that a cluster of galaxies arranged in a line is more likely to occur than a cluster of galaxies arranged in a ring.
\end{abstract}

\keywords{cosmology: theory --- galaxies: clusters: general --- gravitation --- large-scale structure of universe --- methods: analytical --- methods: statistical}

\section{Introduction}

Clusters and groups of galaxies are commonly studied structures in the universe, and have been defined using various criteria; groups are smaller clusters. These clusters may contain over a thousand member galaxies and occupy a few cubic megaparsecs. Although clusters can be identified by their density and size~(e.g. \citealt{1958ApJS....3..211A, 1957PASP...69..409H}) their shapes and structures differ. For example, \citet{1957PASP...69..409H} identify three classes of clusters: \textit{compact clusters} which have a single nearly spherical dense concentration of galaxies, \textit{medium compact clusters} which are less dense and may have multiple concentrations of galaxies and \textit{loose clusters} which do not have any outstanding concentrations of galaxies.

Even these simple classes of clusters suggest greater complexity than just spherical concentrations in a region of space. For example, \citet{1987AJ.....94..251B} found significant substructure in the core of the Virgo cluster as well as pronounced double structure, although \citet{1957PASP...69..409H} classify the Virgo cluster as a medium compact cluster. The structure of the Virgo cluster, as the nearest large cluster, suggests that these irregular shapes are common.

Such irregular shapes may result from mergers. Smaller groups fall into the central region of a cluster and form subgroups whose member galaxies are still tightly bound to each other. Irregular shapes resulting from subgroups then disappear as a cluster virializes. However, many clusters have dynamical relaxation timescales on the order of a Hubble time, and their incomplete virialization suggests that irregular clusters with multiple concentrations of galaxies should be common in the universe.

The basic dynamical description of a cluster is its 6-dimensional phase space configuration such as a sphere with a density and velocity profile. More detailed descriptions of clustering include correlation functions, percolation trees and counts-in-cells statistics. In particular, the counts-in-cells description is especially suitable for this problem because it straightforwardly analyzes regions of space~(cells) with a specified size and shape. In addition, the physics of this description can be derived from gravitational thermodynamics~\citep{1984ApJ...276...13S} or statistical mechanics~\citep{2002ApJ...571..576A} where the basic particles of the system are galaxies that interact in a grand canonical ensemble of cells.

This physical description leads to the gravitational quasi-equilibrium distribution~(GQED) for the counts-in-cells distribution of galaxies. The GQED has been shown to agree with $N$-body simulations~(\citealt{1988ApJ...331...45I} and subsequent work) and various analyses of sky surveys in both the linear and nonlinear regimes of clustering~(e.g. \citealt{2005ApJ...626..795S,2011ApJ...729..123Y} and references therein). This indicates that the statistical mechanical theory is a suitable description of clustering.

The theory was subsequently extended to describe particles with different masses~\citep{2006IJMPD..15.1267A}, three body interactions~\citep{2006ApJ...645..940A} and different internal structures~\citep{2010arXiv1011.0176Y}. It was also further generalized to take into account higher orders in the series expansion of the gravitational interaction~\citep{2010ApJ...720.1246S}, and extended by \citet{2004ApJ...608..636L} to describe the potential and kinetic energies in a single cell, and hence the probability that a cell with $N$ galaxies would be virialized.

In this paper, we extend the work by \citet{2004ApJ...608..636L} and relate the ratio of potential energy to kinetic energy in a cell to its detailed configuration. Here, we work with cells rather than clusters because cells are well-defined regions of space and, unlike clusters, they have a clear boundary. Another advantage of working with cells is that we can use larger cells to study how clusters of galaxies cluster around each other, and hence provide insights into the process of hierarchical structure formation.

This paper is structured as follows: In section \ref{sec-STP} we rescale the thermodynamic variables to describe the properties of a self-gravitating system in a form that is independent of physical units. This provides a scale-free description of particles in a cell. These may be galaxies in a group, or substructures in a cluster, or even clusters in a large cell for insights into hierarchical clustering. In section \ref{sec-PE} we use the scaled thermodynamic variables to determine the probability of finding a cell with a given kinetic energy or virial ratio in quasi-equilibrium. This lets us determine the conditions that produce a virialized or unvirialized cell, and its implications for substructure in these cells. In section \ref{sec-config} we derive a relation between the configuration, energy and probability of a cell. In section \ref{sec-desc} we describe a procedure for studying the internal structure of a cell and apply this procedure to an illustrative example. Finally in section \ref{sec-conc}, we summarize our results.

\section{Scaled Energies}
\label{sec-STP}
We consider spatial cells with a fixed volume and shape, containing $N$ gravitating ``particles'' which for simplicity, have the same mass $m$. We show in Appendix \ref{app-mmass} that if ``particles'' have different masses, the ``particles'' with masses within an order of magnitude of the most massive ``particle'' will dominate the total potential energy.

For these conditions, the scale of a cell and the mass of a ``particle'' are free parameters, and the statistical-mechanical description may apply to very different regimes. ``Particles''  may be individual galaxies in a small cell, tightly bound groups in a larger cell, or even clusters of galaxies in very large cells. Therefore we express the energy of a cell in dimensionless form~\citep{2004ApJ...608..636L} where we write the energy in terms of a scaling factor based on the mass of a ``particle'' and the size of a cell. This lets us focus on the physical properties of clustering and ignore the scale dependent effects.

Because the potential energy of a uniform density cell scales with the mass of a particle and the cell's radius, we use the average potential energy of a particle as a scaling factor. Then the temperature $T$ from the kinetic energy $K$, the correlation potential energy $W$ and total energy $E=W+K$ are expressed in terms of scaled dimensionless quantities~(with Boltzmann's constant unity)
\begin{equation}\label{eq-Tsdef}
T_* \equiv \frac{2K}{3NA},
\end{equation}
\begin{equation}\label{eq-Wsdef}
W_* \equiv \frac{2W}{3NA}
\end{equation}
and
\begin{equation}\label{eq-Esdef}
E_* \equiv \frac{2E}{3NA}
\end{equation}
where $A$ is a scaling factor with dimensions of energy:
\begin{equation}\label{eq-Adef}
A \equiv \frac{3}{4}\frac{N Gm^2}{R_1} \zeta\left(\frac{\epsilon}{R_1}\right)
\end{equation}
and $\zeta(\epsilon/R_1)$ represents the extent of a particle and its halo. Here $R_1$ is the radius over which the configuration integral is taken, which is the distance where the expansion of the universe cancels out the smoothed background potential~\citep{1996ApJ...460...16S}. This is generally the distance at which the two-point correlation function $\xi_2(r)$ becomes negligible.

The scaling factor $A$ is proportional to the absolute value of the average potential energy of a particle, and comes from the potential energy contribution to the partition function~(c.f. equation (14) of \citealt{2002ApJ...571..576A})
\begin{equation}\label{eq-Zconf}
Q_2(T,V) = \int\int\left[1+\frac{Gm^2}{Tr_{12}}\kappa(r,\epsilon)\right]d^3\mathbf{r}_1d^3\mathbf{r}_2.
\end{equation}
Here $G$ is the gravitational constant, $m$ is the mass of a particle. The $\kappa(\epsilon,r)$ and $\zeta(\epsilon/R_1)$ factors are from modifications of the usual point mass potential for extended particles~\citep{2002ApJ...571..576A,2010arXiv1011.0176Y}. The extended length scale $\epsilon$ is typically smaller than $R_1$. Integration by parts gives a relation between $\zeta$ and $\kappa$:
\begin{equation}\label{eq-taudef}
\zeta\left(\frac{\epsilon}{R_1}\right) = \frac{2}{R_1^2}\int_0^{R_1} r\kappa(r, \epsilon)dr.
\end{equation}
In the case of point masses, $\kappa = 1$ and $\zeta = 1$. Except in extreme cases where $\epsilon/R_1$ is very large, $\zeta$ is generally of order unity~(c.f. figure 1 of \citealt{2002ApJ...571..576A}).

The scaled energies given by equations \refeq{eq-Tsdef} and \refeq{eq-Esdef} are essentially ratios of kinetic energy to potential energy, and total energy to potential energy respectively, while the scaled potential energy given by \refeq{eq-Wsdef} describes the ``compactness'' of the spatial configuration. These energies are an instantaneous snapshot of the positions and velocities of galaxies in a cell.

Although these cells are not rigorously in equilibrium, they are in quasi-equilibrium. This means that the instantaneous values of the potential and kinetic energies for a single cell will fluctuate about the current average for an ensemble. On a larger scale, the energies and thermodynamic quantities averaged over an ensemble of cells in quasi-equilibrium change slowly compared to the dynamical timescale within a cell, so intermediate time averages of the ensemble are stable. These fluctuations mean that the instantaneous energies are more more closely related to the local dynamics of a cell than to the thermodynamics of the ensemble. Nevertheless, since the instantaneous energies are distributed about their time-averaged values, they allow us to relate an observed cell to its thermodynamic quantities.

In quasi-equilibrium, the thermodynamic quantities are averages over space and time, and are more closely related to the time-averaged energies of a cell so we use these averages as a basis for our theory. For a cell in quasi-equilibrium, the equation of state for a canonical ensemble~\citep{2002ApJ...571..576A} relates the average total energy of a cell $U$ to its kinetic temperature $T$ and the clustering parameter $b=-W/2K$ through
\begin{equation}\label{eq-Udef}
U = \frac{3NT}{2}(1-2b)
\end{equation}
for units of temperature where the Boltzmann constant is 1.

This means that we can use ensembles with different average energies as an approximation for individual cells with different energies and take $E \approx U$. Under this approximation, the quasi-equilibrium energies are approximately equal to the time-averaged energies. For such ensembles, the clustering parameter $b$ is given by~\citep{2002ApJ...571..576A}
\begin{equation}\label{eq-bdef}
b = -\frac{W}{2K} = \frac{\beta\overline{n}T^{-3}\zeta(\epsilon/R_1)}{1+\beta\overline{n}T^{-3}\zeta(\epsilon/R_1)}
\end{equation}
where $\beta = (3/2)(Gm^2)^3$ and $T$ is the average kinetic temperature of the canonical ensemble.

To relate the scaled energies to the definition of $b$ in equation \refeq{eq-bdef} and to use the thermodynamic quantities described in \citet{2002ApJ...571..576A} for further analysis, we use $R_1 \propto \overline{n}^{-1/3}$ and make the scale transformation following \citet{2002ApJ...571..576A}
\begin{equation}\label{eq-strans}
\frac{G m^2}{T R_1} \to (Gm^2)^3 (R_1 T)^{-3}
\end{equation}
using the scaling property of the partition function derived by \citet[page 93]{1980stph.book.....L}. This transformation avoids fractional powers in $\overline{n}$ and simplifies the derivation of the thermodynamic quantities. Under this transformation we have a different scaling factor $a$ given by
\begin{equation}\label{eq-adef}
a \equiv \frac{3}{2}(Gm^2)^3 \overline{n} \zeta\left(\frac{\epsilon}{R_1}\right)
= \beta \overline{n} \zeta\left(\frac{\epsilon}{R_1}\right)
\end{equation}
and thus define a different set of dimensionless parameters~(c.f. \citealt{2004ApJ...608..636L})
\begin{equation}\label{eq-Tsdefa}
\overline{T}_* \equiv \frac{T}{a^{1/3}} = \frac{2K}{3N a^{1/3}}
\end{equation}
and
\begin{equation}\label{eq-Esdefa}
\overline{E}_* \equiv \frac{2E}{3N a^{1/3}}.
\end{equation}
Using $E \approx U$ and equations \refeq{eq-Udef}, \refeq{eq-bdef} and \refeq{eq-adef} we relate $E_*$ to $T_*$ by
\begin{equation}\label{eq-EsTs1}
\overline{E}_* = \frac{2T(1-2b)}{3N a^{1/3}} = \overline{T}_*\frac{\overline{T}_*^3-1}{\overline{T}_*^3+1}
\end{equation}
which also gives
\begin{equation}\label{eq-Wsdefa}
\overline{W}_* = \frac{2W}{3Na^{1/3}} = \frac{2(E-T)}{3Na^{1/3}} = -\frac{2\overline{T}_*}{\overline{T}_*^3+1}.
\end{equation}
These quantities, derived from the statistical mechanical theory of galaxies interacting in quasi-equilibrium, apply to cells in quasi-equilibrium and represent the time-averaged local energies of a cell taken over the fluctuation timescale of the cell. In contrast, the quantities $T_*$, $W_*$ and $E_*$, are a rescaling of the energies that are associated with a specific phase space configuration, and describe instantaneous values of the energies in a cell.

At this point, we emphasize that the scaled energies $\overline{T}_*$, $\overline{W}_*$ and $\overline{E}_*$, and correspondingly $T_*$, $W_*$ and $E_*$, are local to a cell and do not represent the average of the grand canonical ensemble. In fact, for a system in quasi-equilibrium, the fluctuations in potential energy within a cell are proportional to the local kinetic energy fluctuations so that~\citep{1990ApJ...365..419S,2004ApJ...608..636L}
\begin{equation}\label{eq-alphadef}
GmN\left\langle\frac{1}{r}\right\rangle = \alpha\left\langle v^2 \right\rangle
\end{equation}
where $\langle v^2 \rangle$ is the mean square velocity of galaxies and $\alpha$ is a local form factor that determines the kinetic and potential energy fluctuations. Generally $\alpha$ will vary from volume to volume, and comparing equations \refeq{eq-Tsdef} and \refeq{eq-alphadef}
\begin{equation}\label{eq-Tsalpha}
T_* \propto \frac{1}{\alpha}
\end{equation}
which means that $T_*$ is a local quantity that will vary from cell to cell.

To relate $E_*$ to $\overline{E}_*$, we define the time-averaged local virial ratio $\overline{\psi}$ and its instantaneous counterpart $\psi$ as the ratio of the absolute value of the local correlation potential energy to twice the kinetic energy of galaxies contained within a cell. Hence we get
\begin{equation}\label{eq-Vrdef}
\psi = -\frac{W}{2K} = -\frac{W_*}{2T_*}
\end{equation}
which may fluctuate about its time-averaged value. Using equations \refeq{eq-Wsdefa}, \refeq{eq-Tsdefa} and \refeq{eq-adef}, the corresponding time-averaged value is
\begin{equation}\label{eq-Vrdefa}
\begin{split}
\overline{\psi} &= -\frac{\overline{W}_*}{2\overline{T}_*} = \frac{1}{\overline{T}_*^3+1} = \frac{\overline{T}_*^{-3}}{\overline{T}_*^{-3}+1} \\
&= \left. \frac{\beta\overline{n}T^{-3}\zeta(\epsilon/R_1)}{1+\beta\overline{n}T^{-3}\zeta(\epsilon/R_1)} \right|_{\mathrm{local}}
\end{split}
\end{equation}
which has a range of $0 \leq \overline{\psi} \leq 1$. Although the form of equation \refeq{eq-Vrdefa} is similar to the form of equation \refeq{eq-bdef}, $\overline{\psi}$  and $\psi$ are local quantities that describe the internal virial ratio of a cell whereas $b$ is an average over the entire ensemble. Therefore $\overline{\psi}$ will fluctuate between cells and is distributed with an average value given by $b$.

With the virial ratio, the scaling factors cancel out and the actual observed virial ratio $\psi$ fluctuates about its time-averaged value $\overline{\psi}$. Hence the similarity between equation \refeq{eq-Vrdef} and \refeq{eq-Vrdefa} indicates that we can determine the probability of finding a cell with a given time-averaged energy level, and relate this probability to an actual cluster whose time-averaged energies cannot be observationally determined.

\subsection{Physically important energies}
\label{sec-TsRange}

\begin{figure*}[tbp]
\begin{center}
\parbox{\floatwidth}{\center{(a)\\\includegraphics[width=\floatwidth]{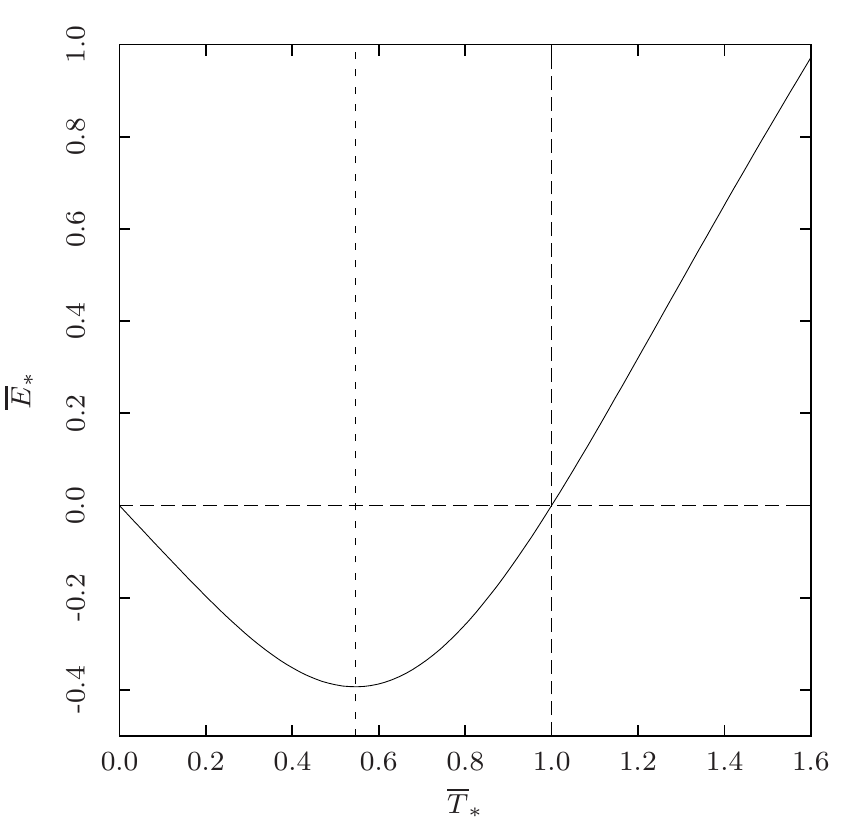}}}
\parbox{\floatwidth}{\center{(b)\\\includegraphics[width=\floatwidth]{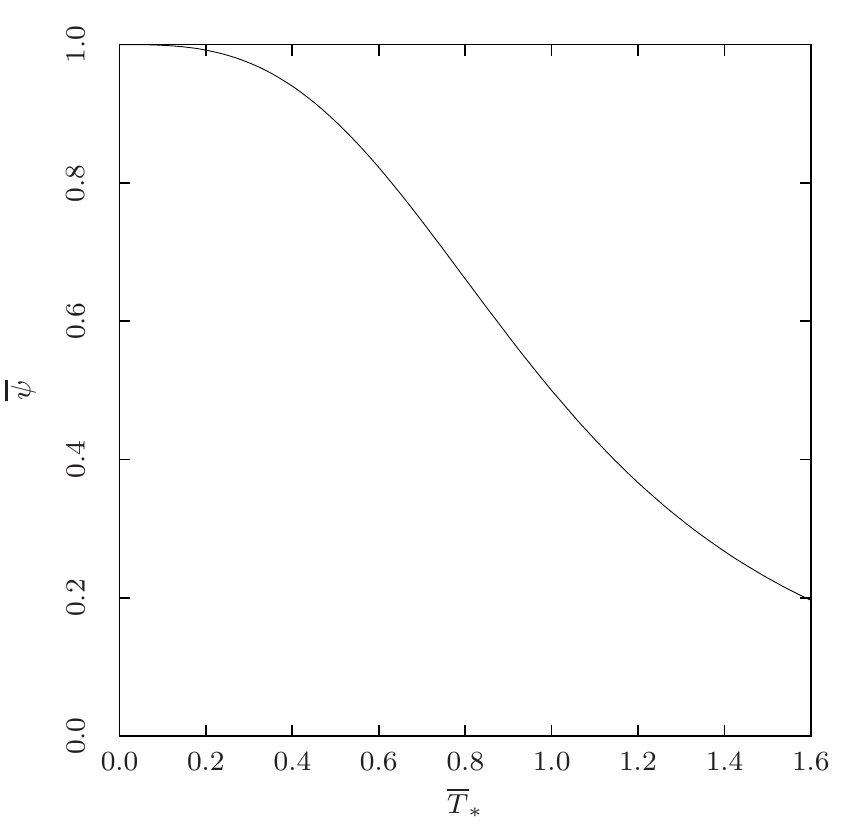}}}
\parbox{\floatwidth}{\center{(c)\\\includegraphics[width=\floatwidth]{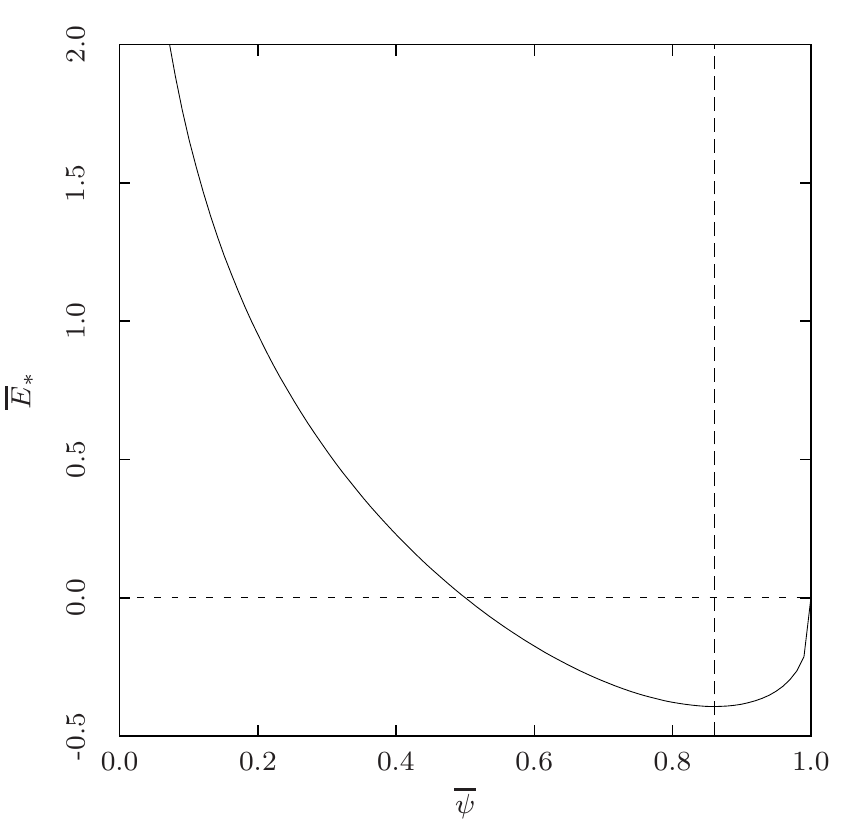}}}
\caption{
(a): The ensemble average normalized energy $\overline{E}_*$, as a function of the ensemble average normalized temperature $\overline{T}_*$. The vertical dotted line separates the region where $\overline{T}_* < 0.54$, corresponding to the negative specific heat branch of $\overline{E}_*[\overline{T}_*]$, from the region where $\overline{T}_* > 0.54$, corresponding to the positive specific heat branch of $\overline{E}_*[\overline{T}_*]$. To the right of the dashed lines, $\overline{E}_* > 0$ and most of the galaxies in the cell are unbound.
(b): The ensemble average normalized virial ratio $\overline{\psi}$, as a function of the ensemble average normalized temperature $\overline{T}_*$. This has the range $0 \leq \overline{\psi} \leq 1$. In comparison, $\overline{T}_*$ may become very large.
(c): The ensemble average normalized energy $\overline{E}_*$, as a function of the ensemble average normalized virial ratio $\overline{\psi}$. $\overline{E}_*$ is negative for $\overline{\psi} > 0.5$ and has a minimum at $\overline{\psi} = (\sqrt{10}-2)^{-1} \approx 0.86$ which is marked by the vertical dashed line.
}
\label{fig-ivar}
\end{center}
\end{figure*}

With the various representations for the energy of a cell, we next examine physically significant values of the time-averaged scaled energies $\overline{E}_*$, $\overline{T}_*$, $\overline{W}_*$ and $\overline{\psi}$ and relations among them. Equation \refeq{eq-EsTs1} indicates that $\overline{T}_*[\overline{E}_*]$ is double valued for $\overline{E}_* \leq 0$ with a minimum at $\overline{T}_* = (\sqrt{10}-3)^{1/3} \approx 0.54$, and is single valued for $\overline{T}_* > 1$. \citet{2004ApJ...608..636L} give a detailed discussion of the different regimes of the $\overline{T}_*[\overline{E}_*]$ relation, so we focus on the  physically relevant cases here. We plot some of the relationships among these quantities in figure \ref{fig-ivar} and indicate some interesting values.

At $\overline{E}_* > 0$, most galaxies in the cell are unbound and the average speed of these galaxies is greater than the escape speed. This corresponds to $\overline{\psi} < 0.5$ and describes the case where clusters are unlikely to form in the cell. Galaxies are mostly bound in clusters when $\overline{E}_* < 0$, corresponding to $\overline{T}_* < 1$, $\overline{\psi} > 0.5$ and $-\overline{W}_* > \overline{T}_*$. At $\overline{E}_* = 0$, $\overline{\psi} = 0.5$, $-\overline{W}_* = \overline{T}_*$ and the kinetic energy is equal to the absolute value of the potential energy.

The state where $\overline{T}_* = 0$ has no average peculiar kinetic energy. This occurs only when the entire universe has collapsed into a single supercluster~\citep{2003MNRAS.345..552B}. For an individual cell, this state is unphysical and hence not in quasi-equilibrium, which suggests that there is a minimum value of $\overline{T}_* > 0$ with a corresponding value of $\overline{E}_*$.

At $\overline{T}_* = (\sqrt{10}-3)^{1/3} \approx 0.54$, $\overline{E}_*[\overline{T}_*]$ is a minimum. This corresponds to $\overline{\psi} = (2+\sqrt{10})/6 \approx 0.86$. We can describe the physical characteristics of this point using the specific heat of the cell, given by
\begin{equation}\label{eq-CVdef}
C_V = \frac{1}{N}\frac{\partial E}{\partial T}
 = \frac{3}{2}\frac{\partial \overline{E}_*}{\partial \overline{T}_*}
 = \frac{3}{2} \frac{\overline{T}_*^6 + 6\overline{T}_*^3 -1}{(1+\overline{T}_*^3)^2}.
\end{equation}
$\overline{T}_* = 0.54$ is a zero of $C_V$ and the specific heat of the cell is negative for $\overline{T}_* < 0.54$ or $\overline{\psi} > 0.86$. Therefore a cell with $\overline{T}_* < 0.54$ is a special case because it is sufficiently condensed that it is unlikely to exchange galaxies and energy with the rest of the universe~\citep{2009arXiv0902.0747S}. Such cells can be approximated by closed systems that continue to relax independently. They form a set of microcanonical ensembles. Other cells, with $\overline{T}_* > 0.54$, may still exchange galaxies and energy with the rest of the universe and thus are part of a grand canonical ensemble. For this reason, we use a grand canonical ensemble for our analysis.

Because the $\overline{T}_*[\overline{E}_*]$ relation is double-valued with a transition at $\overline{T}_* = 0.54$, we describe the branches separately. We refer to the branch where $\overline{T}_* < 0.54$ as the negative specific heat branch since cells with $\overline{T}_* < 0.54$ have negative specific heat, and correspondingly the region where $\overline{T}_* > 0.54$ is the positive specific heat branch.

\subsection{The limits of $\overline{T}_*$}
\label{sec-TsLim}

Because the theory described in this paper is valid for quasi-equilibrium, we examine the conditions needed for a region to be in quasi-equilibrium. The requirement is that the macroscopic evolution of a region in quasi-equilibrium be slow compared with the crossing time of the system so that equilibrium prevails approximately. Regions that are not in quasi-equilibrium are unstable. They generally will restructure internally to reach a quasi-equilibrium state on a timescale that is shorter than the evolution of a quasi-equilibrium system.

The condition that a system be in quasi-equilibrium places limits on $\overline{T}_*$ and $\overline{\psi}$. These limits define the range for the quasi-equilibrium thermodynamic theory to apply. Since we have established that $\overline{T}_* = 0$ is an extreme and unlikely case, we seek physically motivated limits on $\overline{T}_*$. We denote these limits by $\overline{T}_\stx{min}$ and $\overline{T}_\stx{max}$ with corresponding values for $\overline{\psi}$.

In the high temperature limit with large $\overline{T}_*$, the kinetic energy dominates over the potential energy so the system can be described essentially as an ideal~(non-relativistic) gas. This means that systems with large $\overline{T}_*$ are physically possible and can be described simply by the well-studied case of an ideal gas. Therefore $\overline{T}_\stx{max} = \infty$ and $\overline{\psi}_\mathrm{min} = 0$ for the case of an ideal gas.

In the low temperature limit, $\overline{T}_* = 0$ is an extremely unlikely case~(occurring only in static cosmologies) so the quasi-equilibrium lower bound on $\overline{T}_*$ is nonzero. Constraints on $\overline{T}_\stx{min}$ thus have to come from the conditions for quasi-equilibrium for cells on the negative specific heat branch of the $\overline{E}_*[\overline{T}_*]$ relation.

For regions to have negative specific heat, they must be a microcanonical ensemble that has separated out from the grand canonical ensemble that describes the rest of the universe. Because such regions are closed systems, they generally contain clusters that are so tightly bound such that they do not exchange energy or particles with the rest of the universe. This also means that the energy of an individual cluster remains nearly the same as when it first separated out of the grand canonical ensemble. Therefore the quasi-equilibrium grand canonical ensemble can provide initial conditions for the formation of a set of microcanonical ensembles.

To determine the conditions for quasi-equilibrium, we use equation (3.24) of \citet{2004ApJ...608..636L} to relate $\overline{T}_*$ to the mean squared velocity of galaxies in a cell $\langle v^2 \rangle$, the dynamical timescale of the cell $\tau_\mathrm{dyn} \sim (G \overline{\rho})^{-1/2}$ and the average nearest neighbor separation between galaxies $\langle r \rangle \sim \overline{n}^{-1/3}$ by
\begin{equation}\label{eq-Tsvtr}
\overline{T}_* = \frac{\overline{n}^{2/3}\langle v^2 \rangle}{3(3/2)^{1/3}\zeta^{1/3}G\overline{\rho}}
 \approx \frac{0.0881}{\zeta^{1/3}}\frac{\langle v^2 \rangle \tau_\mathrm{dyn}^2}{\langle r \rangle^2}.
\end{equation}

In approximate virial equilibrium, $\langle r \rangle^2/\langle v^2 \rangle \approx \tau_\mathrm{dyn}^2$, which sets the lower bound for $\overline{T}_*$. For $\zeta \approx 1$, this suggests the minimum value of $\overline{T}_*$ is $\overline{T}_\stx{min} \approx 0.1$, corresponding to $\overline{E}_\stx{min}\approx -0.1$ and $\overline{\psi}_\mathrm{max} > 0.999$. With these limits we can now determine the probability that a cell has a given energy. To do so, we use a grand canonical ensemble of cells since cells are essentially open regions of space that are not enclosed by an insulating wall and hence may exchange particles and energy.

\section{Probabilities}
\label{sec-PE}

The probability that a cell with $N$ galaxies in quasi-equilibrium in a grand canonical ensemble has total energy $E$ is given by the usual result from statistical mechanics~(e.g. \citealt{2004ApJ...608..636L})
\begin{equation}\label{eq-pE1}
P(E, N) dE = g(E) \frac{e^{-E/T_0}e^{N\mu/T_0}}{Z_G} dE
\end{equation}
where $g(E)$ is the density of states having energy $E$, and $T_0$, $\mu$ and $Z_G$ are the temperature, chemical potential and partition function of the grand canonical ensemble. Here we use units of temperature where the Boltzmann constant is $1$ so temperature has energy units.

To simplify the analysis, we can separate $E$ and $N$ and express the probability in equation \refeq{eq-pE1} as
\begin{equation}\label{eq-pE1N}
\begin{split}
P_N(E) dE &= f_V(N) P(E|N) dE \\
 &= f_V(N) g(E) \frac{e^{-E/T_0}e^{N\mu/T_0}}{Z_G} dE
\end{split}
\end{equation}
where $P(E|N) dE$ is the conditional probability that a cell has energy $E$ given that it has $N$ galaxies, and $f_V(N)$ is the counts-in-cells distribution given by \citet{2002ApJ...571..576A}. This allows us to work with the subset of cells with the same value of $N$, which projects the grand canonical ensemble into a canonical ensemble.

Using the scaled thermodynamic variables we determine the various factors in equation \refeq{eq-pE1}. From statistical mechanics, the density of states $g(E)$ for an ensemble is
\begin{equation}\label{eq-gE1}
g(E) = \frac{d\Omega}{dE} = \frac{d(e^S)}{dE}
\end{equation}
where $\Omega$ is the number of microstates and $S$ is the entropy of the ensemble. For a canonical ensemble in quasi-equilibrium, the entropy is given by~\citep{2002ApJ...571..576A}
\begin{equation}\label{eq-Sdef}
\begin{split}
S =& -N \ln\left(\frac{N}{VT^{3/2}}\right) + N \ln(1 + \overline{T}_*^{-3}) - \frac{3N\overline{T}_*^{-3}}{1+\overline{T}_*^{-3}}\\
& + \frac{5N}{2} + \frac{3N}{2}\ln\left(\frac{2 \pi m}{\Lambda^2}\right)
\end{split}
\end{equation}
where $T$ is the unscaled kinetic temperature of a cell in units where the Boltzmann constant is 1, $V$ is the volume of a cell, $N$ is the number of galaxies in the cell, $m$ is the mass of a galaxy and $\Lambda$ is a normalizing factor. Since equation \refeq{eq-Esdefa} relates $\overline{E}_*$ to $\overline{T}_*$ we can write the number of microstates $\Omega(\overline{E}_*)$ for scaled energy $\overline{E}_*$ in terms of $\overline{T}_*$ as~(c.f. \citealt{2004ApJ...608..636L})
\begin{equation}\label{eq-OmegaTs}
\Omega(\overline{E}_*[\overline{T}_*]) = \left[\frac{V}{N}\left(1+\overline{T}_*^{-3}\right)\right]^N
 \left(\frac{2 \pi m T}{\Lambda^2}\right)^{\frac{3N}{2}}
 e^{\frac{5N}{2}-\frac{3N}{1+\overline{T}_*^3}}.
\end{equation}
Then $g(\overline{E}_*)$ is
\begin{equation}\label{eq-gE2}
\begin{split}
g(\overline{E}_*[\overline{T}_*]) &= \frac{d\Omega}{d\overline{T}_*}\frac{d\overline{T}_*}{d\overline{E}_*}\\
 &= \frac{3N}{2\overline{T}_*}\left[\frac{V}{N}\left(1+\overline{T}_*^{-3}\right)\right]^N \left(\frac{2 \pi m T}{\Lambda^2}\right)^{\frac{3N}{2}} e^{\frac{5N}{2}-\frac{3N}{1+\overline{T}_*^3}}.
\end{split}
\end{equation}

The other quantities in equation \refeq{eq-pE1} are found in \citet{2002ApJ...571..576A}. For the grand canonical ensemble, the fugacity and partition function are
\begin{equation}\label{eq-nmt}
e^{\frac{N \mu}{T_0}} = \left(\frac{\overline{N}}{V}T_0^{-\frac{3}{2}}\right)^N(1-b)^N e^{-Nb} \left(\frac{2\pi m}{\Lambda^2}\right)^{-\frac{3N}{2}}
\end{equation}
and
\begin{equation}\label{eq-ZG}
Z_G = \exp\left[\overline{N}(1-b)\right]
\end{equation}
where $b$ as in equation \refeq{eq-bdef} is the clustering parameter for the grand canonical ensemble, $\overline{N}$ is the average number of galaxies in a cell, and $T_0$ is the temperature of the grand canonical ensemble. Substituting equations \refeq{eq-gE2}, \refeq{eq-nmt} and \refeq{eq-ZG} into equation \refeq{eq-pE1}, the terms involving $\Lambda$ and $m$ cancel and we get
\begin{equation}\label{eq-pE2}
\begin{split}
&P(\overline{E}_*[\overline{T}_*]|N) d\overline{E}_*\\
 &= \frac{3N}{2\overline{T}_*} \left(\frac{\overline{N}}{N}\left[\frac{T}{T_0}\right]^{3/2}\right)^N (1+\overline{T}_*^{-3})^N (1-b)^N\\
 &\phantom{=}\times\exp\left[\frac{5N}{2} - \frac{3N}{1+\overline{T}_*^3} - Nb -\overline{N}(1-b)-\frac{E}{T_0}\right] d\overline{E}_*.
\end{split}
\end{equation}

To write equation \refeq{eq-pE2} in terms of $\overline{T}_*$ we solve for $T/T_0$ and $E/T_0$. Equations \refeq{eq-adef} and \refeq{eq-bdef} give
\begin{equation}\label{eq-bT0}
T_0 = \left[\frac{1-b}{b} \beta\overline{n}\zeta(\epsilon/R_1)\right]^{1/3}
 = \left[\frac{1-b}{b} a\right]^{1/3}.
\end{equation}
Using equation \refeq{eq-Tsdefa} we get
\begin{equation}\label{eq-TsT0}
\frac{T}{T_0} = \overline{T}_*\left[\frac{b}{1-b}\right]^{1/3}
\end{equation}
and from equation \refeq{eq-Esdefa} we get
\begin{equation}\label{eq-EsT0}
\frac{E}{T_0} = \frac{3N}{2}\overline{T}_*\frac{\overline{T}_*^3-1}{\overline{T}_*^3+1}\left[\frac{b}{1-b}\right]^{1/3}.
\end{equation}
Substituting equations \refeq{eq-TsT0} and \refeq{eq-EsT0} into equation \refeq{eq-pE2}, we get
\begin{equation}\label{eq-pE3}
\begin{split}
&P(\overline{E}_*[\overline{T}_*]|N) d\overline{E}_* \\
&= \frac{3N}{2\overline{T}_*}\left(\frac{\overline{N}}{N}\sqrt{\frac{\overline{T}_*^3 b}{1-b}}\right)^N (1+\overline{T}_*^{-3})^N (1-b)^N \\
 &\phantom{=}\times\exp\left[(N - \overline{N})(1-b) + \frac{3N}{2}\frac{\overline{T}_*^3-1}{\overline{T}_*^3+1} \right.\\
 &\phantom{=}\times\left.\left(1-\left[\frac{\overline{T}_*^3 b}{1-b}\right]^{\frac{1}{3}}\right)\right] d\overline{E}_*
\end{split}
\end{equation}
which gives the differential conditional probability that a cell has scaled energy $\overline{E}_*$ given that it has $N$ galaxies.

\subsection{Change of Variables and Normalization of $P(\overline{E}_*[\overline{T}_*]|N)$}

The probability that a cell has a scaled energy in the range $\overline{E}_\stx{1} \leq \overline{E}_* \leq \overline{E}_\stx{2}$ comes from integrating over the relevant range so that
\begin{equation}\label{eq-PE}
P_N(\overline{E}_\stx{1} \leq \overline{E}_* \leq \overline{E}_\stx{2}) = \int_{\overline{E}_\stx{1}}^{\overline{E}_\stx{2}} P(\overline{E}_*[\overline{T}_*]|N)d\overline{E}_*
\end{equation}
which is normalized by integrating over all possible values of $\overline{E}_*$
\begin{equation}\label{eq-PENorm}
P_{N,\mathrm{norm}} = \int_{\overline{E}_\stx{min}}^{\overline{E}_\stx{max}} P(\overline{E}_*[\overline{T}_*]|N)d\overline{E}_*.
\end{equation}
Because $\overline{E}_*[\overline{T}_*]$ has a double valued regime for $\overline{E}_* < 0$, the integrals in equations \refeq{eq-PE} and \refeq{eq-PENorm} are taken by integrating over both the positive and negative specific heat branches. We can split the integral such that
\begin{equation}\label{eq-PESep}
\begin{split}
&\int_{\overline{E}_\stx{1}}^{\overline{E}_\stx{2}} P(\overline{E}_*[\overline{T}_*]|N)d\overline{E}_* \\
&= \int_{\overline{E}_\stx{1}}^{\overline{E}_\stx{2}} P_-(\overline{E}_*[\overline{T}_*]|N)d\overline{E}_* + \int_{\overline{E}_\stx{1}}^{\overline{E}_\stx{2}} P_+(\overline{E}_*[\overline{T}_*]|N)d\overline{E}_*
\end{split}
\end{equation}
where $P_-$ and $P_+$ denote probabilities for the negative and positive specific heat branches of $\overline{T}_*[\overline{E}_*]$ respectively. These ranges are also subject to the quasi-equilibrium limits such that $\overline{T}_* > 0.1$ and $\overline{E}_* \geq -0.390$.

To simplify the analysis, we can rewrite the probability in terms of $\overline{T}_*$ and $\overline{\psi}$. The change of variables to rewrite the probability in terms of $\overline{T}_*$ is
\begin{equation}\label{eq-pT1}
P(\overline{T}_*|N) dT_* = P(\overline{E}_*[\overline{T}_*]|N) \left| \frac{d\overline{E}_*}{d\overline{T}_*}\right| d\overline{T}_*
\end{equation}
where we take the absolute value of the Jacobian
\begin{equation}\label{eq-pT1J}
\frac{d\overline{E}_*}{d\overline{T}_*} = \frac{\overline{T}_*^6 + 6 \overline{T}_*^3 -1}{(1+\overline{T}_*^3)^2}
\end{equation}
because it is negative in the negative specific heat branch where $\overline{T}_* < 0.54$. The probability is therefore
\begin{equation}\label{eq-pT}
\begin{split}
P(\overline{T}_*|N) dT_* =& \overline{N}^Ne^{-\overline{N}(1-b)} \left|\frac{\overline{T}_*^6 + 6 \overline{T}_*^3 -1}{(1+\overline{T}_*^3)^2}\right|\frac{3N}{2\overline{T}_*} \\
&\times \left[\frac{1}{N}\sqrt{\frac{\overline{T}_*^3 b}{1-b}}(1+\overline{T}_*^{-3})(1-b)\right]^N\\
&\times \exp\left[N(1-b) + \frac{3N}{2}\frac{\overline{T}_*^3-1}{\overline{T}_*^3+1}\right.\\
&\left.\times\left(1-\left[\frac{\overline{T}_*^3 b}{1-b}\right]^{1/3}\right)\right] d\overline{T}_*
\end{split}
\end{equation}
where we have factored out terms containing $\overline{N}$. These terms will cancel out when the probability is normalized because they do not depend on $\overline{T}_*$. The integral in equation \refeq{eq-PESep} becomes
\begin{equation}\label{eq-PTAbs}
P_N(\overline{T}_\stx{1} \leq \overline{T}_* \leq \overline{T}_\stx{2}) = \int_{\overline{T}_\stx{1}}^{\overline{T}_\stx{2}} P(\overline{T}_*|N)d\overline{T}_*
\end{equation}
which includes both the positive and negative specific heat branches. The normalization factor follows as
\begin{equation}\label{eq-PTNorm}
P_{N,\mathrm{norm}} = \int_{\overline{T}_\stx{min}}^{\overline{T}_\stx{max}} P(\overline{T}_*|N) d\overline{T}_*
\end{equation}
which is simpler to evaluate because $\overline{T}_*$ is single-valued throughout the domain of integration.

Similarly to equation \refeq{eq-pT1}, we write the probability in terms of $\overline{\psi}$ as
\begin{equation}\label{eq-ppsi1}
\begin{split}
P(\overline{\psi}|N) d\overline{\psi} &= P(\overline{T}_*|N)\left|\frac{d\overline{T}_*}{d\overline{\psi}}\right| d\overline{\psi} \\
 &= P(\overline{E}_*|N)\left|\frac{d\overline{E}_*}{d\overline{T}_*}\frac{d\overline{T}_*}{d\overline{\psi}}\right| d\overline{\psi}
\end{split}
\end{equation}
where we take the absolute value of the Jacobian
\begin{equation}\label{eq-ppsi1J}
\frac{d\overline{E}_*}{d\overline{\psi}} = \frac{1+4 \overline{\psi}-6 \overline{\psi}^2}{3(1-\overline{\psi})^{2/3} \overline{\psi}^{4/3}}
\end{equation}
because it is negative for the negative specific heat branch. Using equation \refeq{eq-Vrdef} we write the probability in terms of $\overline{\psi}$ to get
\begin{equation}\label{eq-ppsi}
\begin{split}
P(\overline{\psi}|N) d\overline{\psi} =& \overline{N}^Ne^{-\overline{N}(1-b)}\left|\frac{1+4 \overline{\psi}-6 \overline{\psi}^2} {3(1-\overline{\psi})^{2/3} \overline{\psi}^{4/3}}\right| \frac{3N}{2}\\
&\times \left(\frac{\overline{\psi}}{1-\overline{\psi}}\right)^{1/3} \left[\frac{1}{N}\sqrt{\frac{(1-b) b}{(1-\overline{\psi})\overline{\psi}}}\right]^N \\
& \times \exp\left[N(1-b) + \frac{3N}{2}(1-2 \overline{\psi})\right.\\
&\left.\times \left(1-\left[\frac{(1-\overline{\psi})b} {\overline{\psi}(1-b)}\right]^{1/3}\right)\right] d \overline{\psi}.
\end{split}
\end{equation}
The probability that a cell has a virial ratio in the range $\overline{\psi}_1 \leq \overline{\psi} \leq \overline{\psi}_2$ is thus
\begin{equation}\label{eq-PpsiAbs}
P_N(\overline{\psi}_1 \leq \overline{\psi} \leq \overline{\psi}_2) = \int_{\overline{\psi}_1}^{\overline{\psi}_2} P(\overline{\psi}|N)d\overline{\psi}
\end{equation}
which is normalized by
\begin{equation}\label{eq-PpsiNorm}
P_{N,\mathrm{norm}} = \int_{\overline{\psi}_\mathrm{min}}^{\overline{\psi}_\mathrm{max}} P(\overline{\psi}|N) d\overline{\psi}.
\end{equation}

\begin{figure*}[tbp]
\begin{center}
\includegraphics[width=\floatwidth]{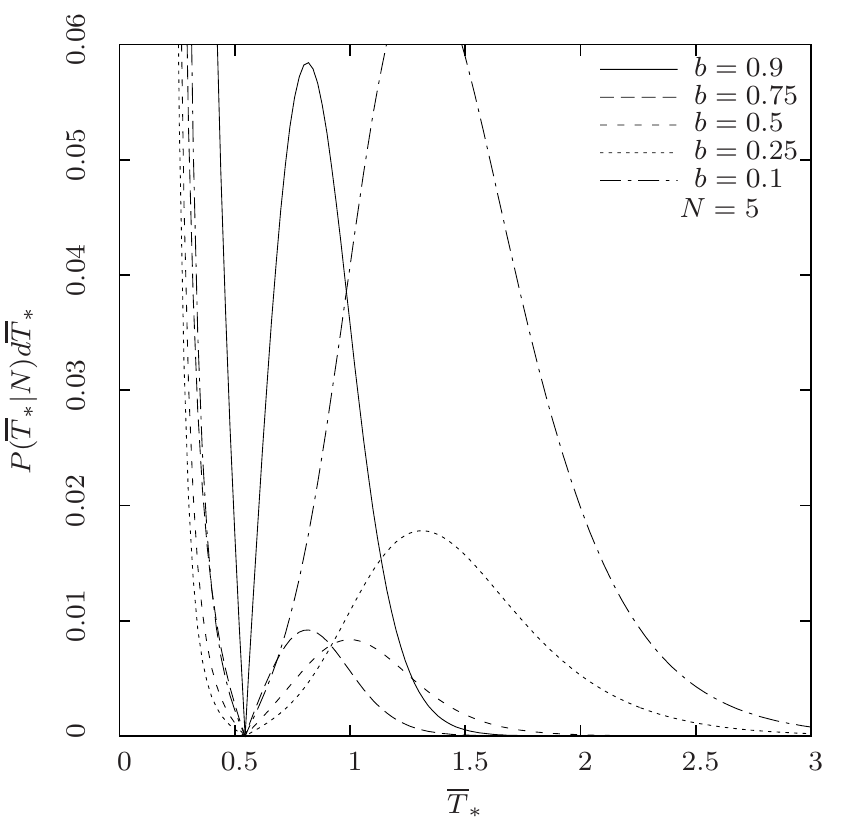}
\includegraphics[width=\floatwidth]{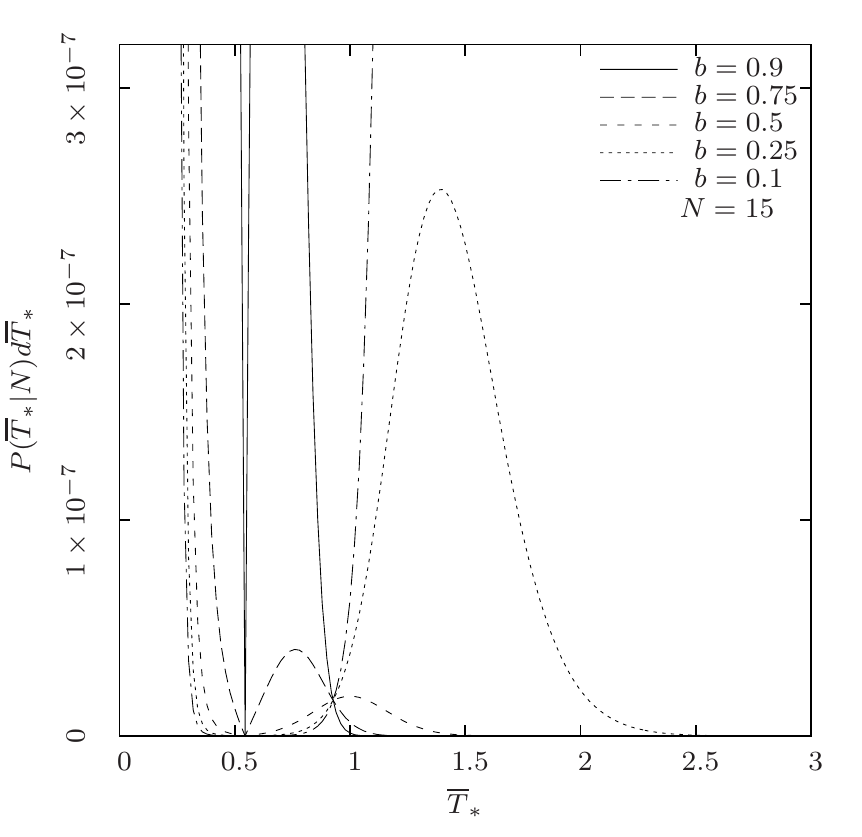}\\
\includegraphics[width=\floatwidth]{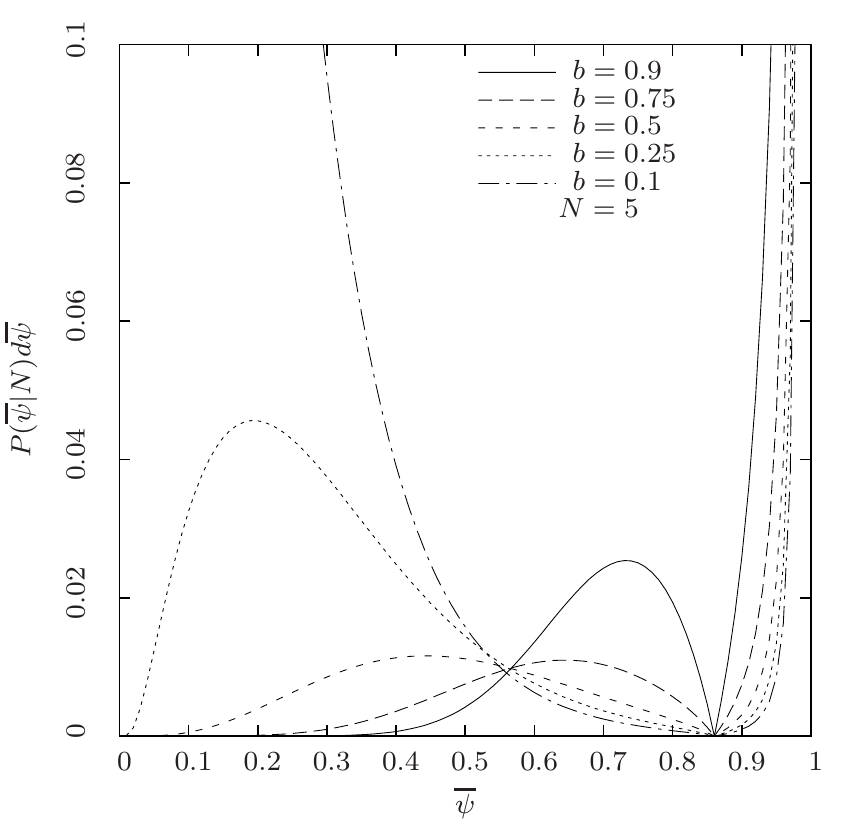}
\includegraphics[width=\floatwidth]{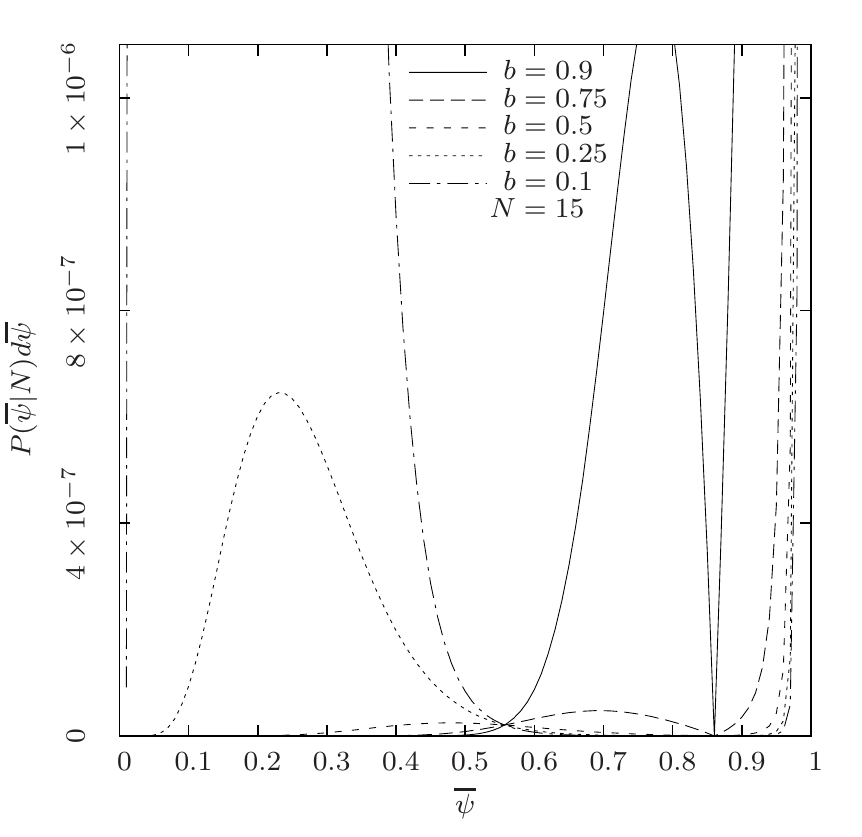}
\caption{
Top row: $P(\overline{T}_*)$ for different values of $N$ and $b$ plotted as a function of $\overline{T}_*$. The left panel is for $N = 5$ and the right panel is for $N = 15$.
Bottom row: $P(\overline{\psi})$ for different values of $N$ and $b$ plotted as a function of $\overline{\psi}$. The left panel is for $N = 5$ and the right panel is for $N = 15$. All plots are independent of $\overline{N}$ because $\overline{N}$ cancels out in the normalization.
}
\label{fig-prob}
\end{center}
\end{figure*}

Here the $\overline{T}_*$ and $\overline{\psi}$ representations~(equations \ref{eq-pT} and \ref{eq-ppsi}) of the probability essentially describe the same system, but provide insights into different physical properties. Using $\overline{T}_\stx{min} = 0.1$ and $\overline{T}_\stx{max} = \infty$ as discussed in the previous section, we normalize and plot $P(\overline{T}_*|N)$ and $P(\overline{\psi}|N)$ in figure \ref{fig-prob}.

Although the unnormalized probabilities given in equations \refeq{eq-pT} and \refeq{eq-ppsi} contain factors involving $\overline{N}$, these factors of $\overline{N}$ are independent of $\overline{T}_*$ and $\overline{\psi}$ and thus cancel out in the normalization. Therefore with the normalization given by equations \refeq{eq-PTNorm} and \refeq{eq-PpsiNorm}, the probabilities in equations \refeq{eq-PTAbs} and \refeq{eq-PpsiAbs} are independent of $\overline{N}$.

From figure \ref{fig-prob}, the probabilities are mostly concentrated in the negative specific heat branch, suggesting that galaxies in a cell are very likely to form a cluster that separates out into a microcanonical ensemble. This shows up especially in the case for $N = 15$ where the normalized probabilities for the positive specific heat branch are on the order of $10^{-8}$. In general, denser cells with larger $N$ are more likely to have bound and collapsing clusters because the more galaxies there are in a cell, the more likely they are to form a cluster.

\subsection{The probability of a bound cluster}
From the probability that a cell has a given energy, we can calculate the probability that its galaxies are bound, and the probability that it has a negative specific heat. In these cells, the absolute value of their internal potential energy is greater than their kinetic energy so that most galaxies do not have the escape speed. Thus $\overline{E}_* < 0$ which corresponds to $\overline{T}_* < 1$. The probability that a cell is bound is therefore
\begin{equation}\label{eq-pbound}
P_\mathrm{bound}(N, b) = \frac{\int_{\overline{T}_\stx{min}}^{1}P(\overline{T}_*|N)d\overline{T}_*}
  {\int_{\overline{T}_\stx{min}}^{\infty} P(\overline{T}_*|N)d\overline{T}_*}
\end{equation}
and the probability that a cell has negative specific heat is
\begin{equation}\label{eq-pneg}
P_\mathrm{neg}(N, b) = \frac{\int_{\overline{T}_\stx{min}}^{0.54}P(\overline{T}_*|N)d\overline{T}_*}
  {\int_{\overline{T}_\stx{min}}^{\infty} P(\overline{T}_*|N)d\overline{T}_*}.
\end{equation}
The probability that a cell has negative specific heat given that it is bound is similarly calculated from
\begin{equation}\label{eq-pbneg}
P_\mathrm{neg|bound}(N, b) = \frac{\int_{\overline{T}_\stx{min}}^{0.54}P(\overline{T}_*|N)d\overline{T}_*}
  {\int_{\overline{T}_\stx{min}}^{1} P(\overline{T}_*|N)d\overline{T}_*}.
\end{equation}

\begin{figure*}[tbp]
\begin{center}
\parbox{\floatwidth}{\center{(a)\\\includegraphics[width=\floatwidth]{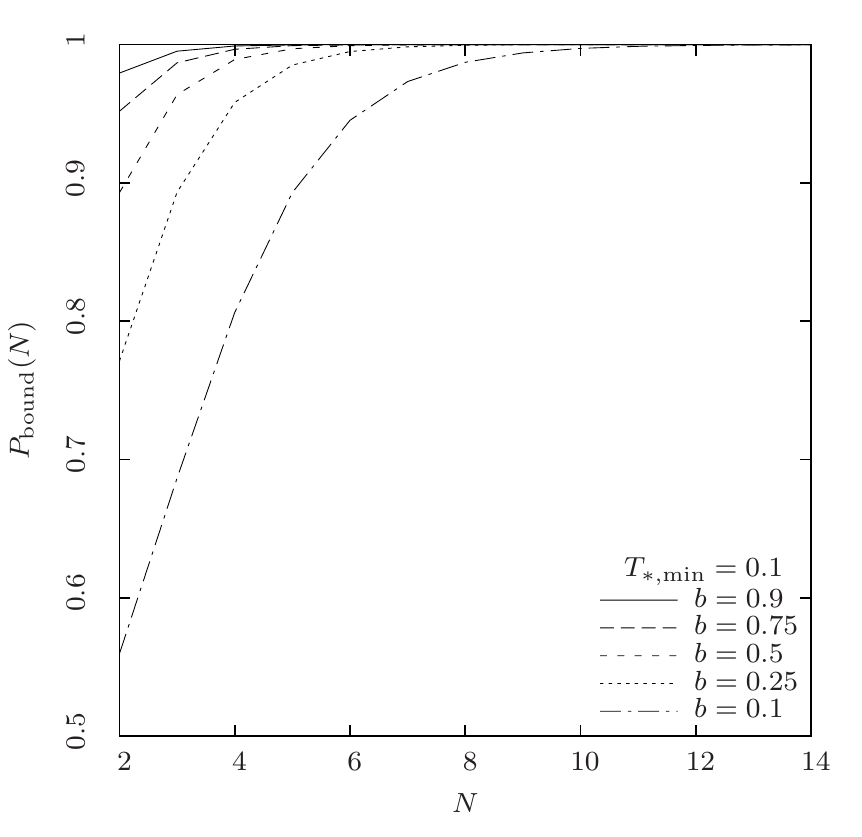}}}
\parbox{\floatwidth}{\center{(b)\\\includegraphics[width=\floatwidth]{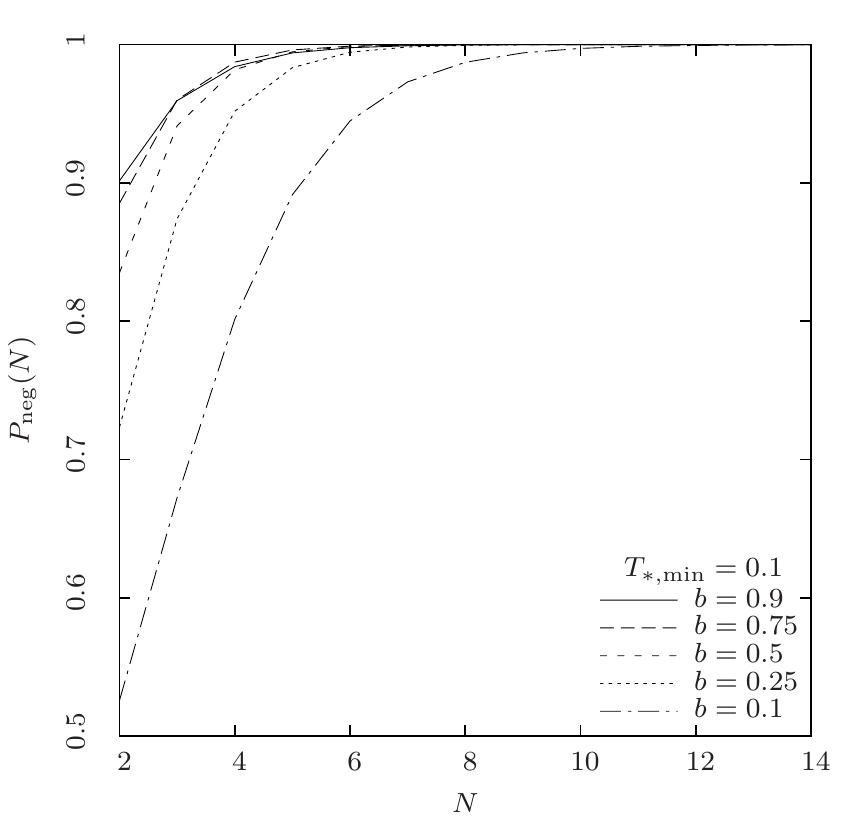}}}
\parbox{\floatwidth}{\center{(c)\\\includegraphics[width=\floatwidth]{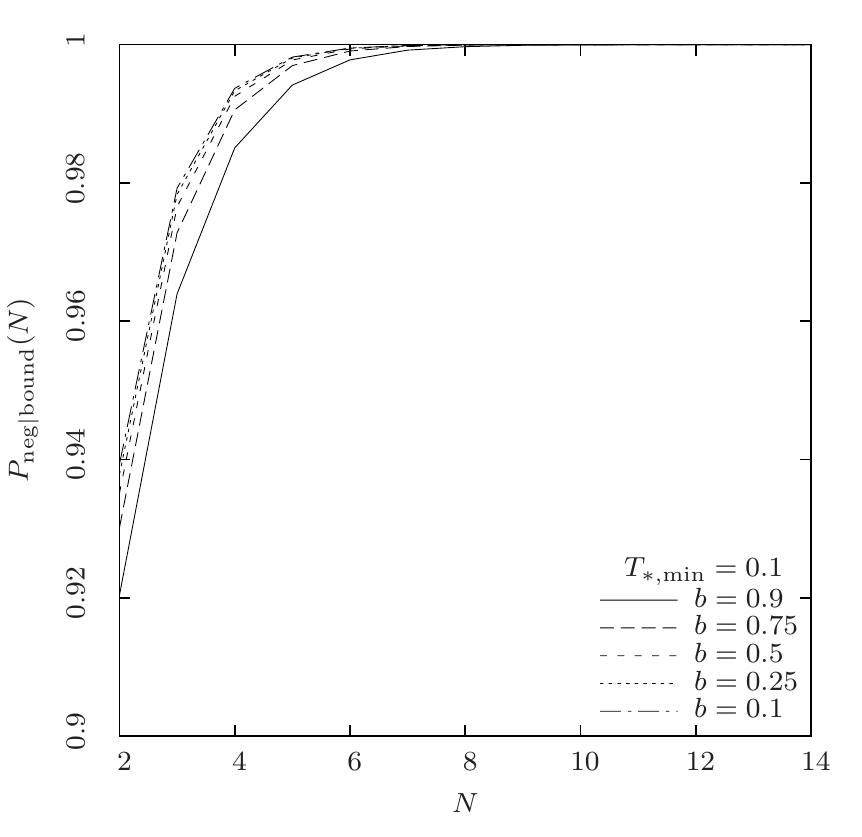}}}
\caption{
Probabilities for different values of $b$ plotted as a function of $N$ using $\overline{T}_\stx{min} = 0.1$.
(a): Probability that galaxies in a cell are bound $P_\mathrm{bound}$.
(b): Probability that a cell has negative specific heat $P_\mathrm{neg}$.
(c): Probability that a cell has negative specific heat given that it is bound $P_\mathrm{neg|bound}$.
}
\label{fig-pbound}
\end{center}
\end{figure*}

We use $\overline{T}_\stx{min} = 0.1$ as discussed in section \ref{sec-TsLim} and plot these probabilities against $N$ for different values of $b$ in figure \ref{fig-pbound}. For dense cells with $N$ greater than about 10, the galaxies in a cell are likely to be bound in a cluster with negative specific heat. This suggests that such clusters may already have collapsed into a microcanonical ensemble and their internal energies are mostly inaccessible to the rest of the universe. Such clusters are also more likely to be relaxed than clusters with fewer galaxies because they are denser and have a faster relaxation timescale. Unlike $P_{\mathrm{neg}}$ and $P_{\mathrm{bound}}$, $P_{\mathrm{neg|bound}}$ decreases as $b$ increases. This is because as $b$ increases, cells are relatively more likely to be bound than they are to have negative specific heat.

\section{Energy and Shape}
\label{sec-config}
Having determined the probability that a cell has a particular energy, the next step is to relate energy to shape. The most complete description of shape gives the detailed positions of all the galaxies in configuration space which involves $6N$ independent variables. We can, however, take advantage of a number of properties of the galaxy distribution to simplify the problem.

First, the mass dependence is contained in the factor $\beta = (3/2)(G m^2)^3$ which scales to describe the quasi-equilibrium clustering of a system in which all objects have the same mass. This allows us to describe a hierarchical model of clustering with different mass ``particles'' at each level of the hierarchy by rescaling $\beta$.

Next, because halos, galaxies and clusters of galaxies that are mostly relaxed and have a virial ratio greater than about $0.86$ have a negative specific heat, they approximately form a set of microcanonical ensembles. Hence their internal energies are largely inaccessible to the rest of the universe. In particular, small dense groups are more likely to be virialized than larger clusters because their dynamical timescales, given by $\tau_\mathrm{dyn} \propto (G \rho)^{-1/2}$ are shorter. This allows us to describe virialized groups of galaxies as individual particles that can be treated as modified point masses. Finally, merging particles that are bound to each other can be treated as a single extended particle~\citep{2010arXiv1011.0176Y}. This allows us to describe merging objects in the statistical mechanical theory.

These properties mean that we can represent many clusters of galaxies as groups of mostly virialized subclusters. The shapes of such clusters are thus given by the positions of their subclusters which reduces the many-body problem of describing all the galaxies in a cluster to a few-body problem that describes how subclusters interact with each other. Based on the probability that a cell is not completely bound and virialized as discussed in the previous section, we expect that most cells will usually contain fewer than about 10 of these subclusters, depending on the value of $b$. Some cells, however, can be dominated by a single virialized cluster.

\subsection{The Potential Energy of Subclusters}

Subclusters can be described by the positions of their individual galaxies with respect to the subcluster's center-of-mass and the positions of these centers-of-mass. To illustrate this, we use an example of a cluster composed of two subclusters. The total potential energy of such a cluster is given by the sum of the internal potential energies of its subclusters and their mutual potential energies such that
\begin{equation}\label{eq-PEBinary}
\begin{split}
\Phi &= -G\left[
   \frac{1}{2}\sum_{i\neq j}\frac{m_i^{(1)}m_j^{(1)}} {\left|\mathbf{x}_i^{(1)}-\mathbf{x}_j^{(1)}\right|}
 + \frac{1}{2}\sum_{i\neq j}\frac{m_i^{(2)}m_j^{(2)}} {\left|\mathbf{x}_i^{(2)}-\mathbf{x}_j^{(2)}\right|} \right.\\
&\phantom{=}\left. + \sum_{i^{(1)}, j^{(2)}}\frac{m_i^{(1)}m_j^{(2)}} {\left|\mathbf{x}_i^{(1)}-\mathbf{x}_j^{(2)}\right|}
\right] \\
 &= \Phi^{(1)} + \Phi^{(2)} -\sum_{i^{(1)}, j^{(2)}}
    \frac{G m_i^{(1)}m_j^{(2)}}{\left|\mathbf{x}_i^{(1)}-\mathbf{x}_j^{(2)}\right|}
\end{split}
\end{equation}
where the superscripts indicate particles from different subclusters. The first two terms are the internal potential energies of the subclusters, which are inaccessible to the rest of the ensemble if these subclusters are virialized. The last term represents the potential energy between the subclusters. We can write this term as
\begin{equation}\label{eq-PEBCross}
\begin{split}
\Phi^{(1,2)}
 &= -\sum_{i^{(1)}, j^{(2)}}\frac{G m_i^{(1)}m_j^{(2)}}{\left|\mathbf{x}_i^{(1)}-\mathbf{x}_j^{(2)}\right|}\\
 &= -\sum_{i^{(1)}, j^{(2)}}\frac{G m_i^{(1)}m_j^{(2)}}{r^{(1,2)}}\frac{r^{(1,2)}}{\left|\mathbf{x}_i^{(1)}-\mathbf{x}_j^{(2)}\right|}
\end{split}
\end{equation}
where $r^{(1,2)}$ is the separation between the centers-of-mass of each subcluster. The detailed internal structure of the subclusters modifies their interaction potential. This modification is
\begin{equation}\label{eq-tauavg}
\kappa(r^{(1,2)}) = \frac{1}{M^{(1)} M^{(2)}}
  \sum_{i^{(1)}, j^{(2)}} \frac{m_i^{(1)} m_j^{(2)}r^{(1,2)}}
  {\left|\mathbf{x}_i^{(1)}-\mathbf{x}_j^{(2)}\right|}
\end{equation}
where $M^{(1)}$ and $M^{(2)}$ are the masses of subclusters 1 and 2 respectively. Then the potential between subclusters is
\begin{equation}\label{eq-PEBCross2}
\Phi^{(1,2)} = -\frac{G M^{(1)} M^{(2)}}{r^{(1,2)}}\kappa(r^{(1,2)})
\end{equation}
where $\kappa(r^{(1,2)})$ describes the modification to the point-mass potential by a pair of extended structures. This modification term enters as a coefficient to $\beta$~\citep{2002ApJ...571..576A,2010arXiv1011.0176Y} and if $\kappa(r^{(1,2)})$ is close to unity its effect will be small.

We can write the positions of galaxies as the vector sum of the position of their subcluster's center-of-mass and their position within the subcluster:
\begin{equation}\label{eq-subcpos}
\mathbf{x}_i^{(1)} = \mathbf{x}^{(1)} + \tilde{\mathbf{x}}_i^{(1)}
\end{equation}
where $\mathbf{x}^{(1)}$ is the center-of-mass of subcluster 1 and $\tilde{\mathbf{x}}_i^{(1)}$ is the position of particle $i$ with respect to the center-of-mass of subcluster 1. With this notation, the modification term in equation \refeq{eq-PEBCross} has the limits
\begin{equation}\label{eq-tauCondition}
1 \geq
\frac{r^{(1,2)}}{\left|\mathbf{x}_i^{(1)}-\mathbf{x}_j^{(2)}\right|} \geq \frac{r^{(1,2)}}
{r^{(1,2)} + \left|\tilde{\mathbf{x}}_i^{(1)} - \tilde{\mathbf{x}}_j^{(2)}\right|}.
\end{equation}

In the limit where $|\tilde{\mathbf{x}}_i^{(1)} - \tilde{\mathbf{x}}_j^{(2)}| \ll r^{(1,2)}$ the modification term is 1. This means that we can approximate subclusters that are widely separated as point masses when computing their interaction potential. For subclusters that are touching, $\kappa(r^{(1,2)}) \approx 0.5$ because $|\tilde{\mathbf{x}}_i^{(1)}-\tilde{\mathbf{x}}_j^{(2)}|$ is on average the radius of a subcluster. Subclusters that are closer to each other may have a smaller value of $\kappa(r^{(1,2)})$, but such pairs may be merging, in which case we can treat such pairs as single subclusters with a different internal structure~\citep{2010arXiv1011.0176Y}.

This means that the many-body problem of studying all the galaxies in a cluster is reduced to the few-body problem of studying the positions and velocities of the subclusters in the cluster. This makes the problem considerably easier, since there are fewer ``particles'' to deal with. Because clusters with more than 10 particles are likely to be virialized, we consider the case where cells have less than 10 subclusters. These cells have a non-negligible probability of having a positive specific heat.

Although subclusters may have different masses and different internal structures, there are analyses that take into account these more general cases. The analysis for particles of different internal structure is described in \citet{2010arXiv1011.0176Y}, and as suggested by equation \refeq{eq-tauavg} enters as a coefficient to $\beta$. The analysis for multiple masses is considerably more complicated~\citep{2006IJMPD..15.1267A}, so to simplify it we make the reasonable approximation that ``particles'' have the same mass. In Appendix \ref{app-mmass} we show that the masses of individual particles are typically within an order of magnitude of each other.

\subsection{The Detailed Configuration of a Cell}
The detailed configuration of a cell is directly related to its energy. We consider its potential and kinetic energies separately, and relate them to $W_*$ and $T_*$. We work with the instantaneous values of the energies since these quantities are well-defined and can be determined in principle by taking a snapshot of a cluster at a given time. In this section, we refer to the basic object in a cell as a ``particle'', bearing in mind that it may describe individual galaxies or virialized subclusters.

\subsubsection{Potential Energy}
The potential energy in the cell can be separated into a local and background component such that
\begin{equation}\label{eq-Phitotal}
\Phi = \Phi_\mathrm{local} + \Phi_\mathrm{background}
\end{equation}
where $\Phi_\mathrm{local}$ is the potential energy that arises from pairs where both members of the pair are within the cell and $\Phi_\mathrm{background}$ is the potential energy that results from pairs that have one member in the cell and the other member outside the cell.

The local potential energy in the cell comes from summing up the potential energy of all pairs within the cell such that
\begin{equation}\label{eq-Philocaldef}
\Phi_\mathrm{local} = -\sum_{1\leq i < j \leq N} \frac{Gm^2}{r_{ij}}\kappa(r_{ij})
\end{equation}
where $G$ is the gravitational constant, $m$ is the mass of a galaxy, $r_{ij}$ is the distance between particle $i$ and particle $j$ and the sum is taken over all possible pairs of galaxies within the cell. Here $\kappa(r)$ describes the potential of particles that are extended sources and is a modification to the potential used to describe point masses. It is related to $\zeta(\epsilon/R_1)$ through equation \refeq{eq-taudef} and is usually of order unity for particles that are small compared to the cell size.

We can write the local potential in terms of the average inverse separation of galaxy pairs so that
\begin{equation}\label{eq-Phiavg}
\Phi_\mathrm{local} = -G m^2 \frac{N(N-1)}{2} \left\langle \frac{\kappa(r)}{r}\right\rangle
\end{equation}
where the $N(N-1)/2$ factor is the number of unique pairs in a cell with $N$ particles. Here $\langle 1/r\rangle$ is the average inverse pairwise separation of all galaxy pairs in the cell.

In an expanding universe, the expansion of the universe exactly cancels the smoothed background term in the potential~\citep{1996ApJ...460...16S} so we can ignore the background term for cells that are larger than the scale at which the two-point correlation function $\xi_2$ is negligible. At these scales, the potential energy is extensive since the expansion of the universe cancels out the background term. For smaller cells, \citet{1996ApJ...460...16S} suggest that extensivity is also a good approximation in the regime where $\xi_2 \gtrsim 1$ so that the correlation energy within a cell is much greater than the correlation energy between cells. This approximation holds because the form of the partition function is independent of scale. The scale dependence is provided by the clustering parameter $b$ which is related to the two-point correlation function $\xi_2$ through~\citep{1996ApJ...460...16S}
\begin{equation} \label{eq-bcorr}
b \equiv \frac{2 \pi G m^2 \overline{n}}{3T} \int_0^R \xi_2(r) r dr.
\end{equation}

For these reasons, the potential energy in a cell is approximately extensive regardless of its radius, and the local potential $\Phi_\mathrm{local}$ is a good approximation for the correlation potential energy. This also lets us use $R$ instead of $R_1$ in the scaling factors $A$ and $a$

For an individual cell, the potential energy, $W$, is thus
\begin{equation}\label{eq-Wdef2}
W = \Phi_\mathrm{local} = -G m^2 \frac{N(N-1)}{2} \left\langle\frac{\kappa(r)}{r}\right\rangle
\end{equation}
and using equation \refeq{eq-Wsdef}, we can write the scaled potential energy as
\begin{equation}\label{eq-Wsdef2}
\begin{split}
W_* &= -\frac{2 G m^2}{3NA}\frac{N(N-1)}{2} \left\langle\frac{\kappa(r)}{r}\right\rangle \\
 &= -\frac{4}{9}\frac{(N-1)}{N\zeta(\epsilon/R)}R\left\langle\frac{\kappa(r)}{r}\right\rangle
\end{split}
\end{equation}
where we have used the scaling factor $A$ given by equation \refeq{eq-Adef} because we are dealing with the detailed configuration of the galaxies in the cell. Here, the $Gm^2$ factors cancel and $W_*$ is determined only by $N$, $R$ and $\langle \kappa(r)/r\rangle$.

For a homogeneous cell, the average inverse separation of all pairs of particles scaled to the cell radius is
\begin{equation}\label{eq-ravgH}
R\left\langle \frac{1}{r}\right\rangle_\mathrm{hom.}
 = \frac{R}{N} \int_0^R \frac{N}{V} \frac{1}{r} 4 \pi r^2 dr = \frac{3}{2}.
\end{equation}
We can then write
\begin{equation}\label{eq-Ravgform}
\frac{N-1}{N}R\left\langle\frac{\kappa(r)}{r}\right\rangle
 = R\left\langle\frac{1}{r}\right\rangle_\mathrm{hom.} \eta  = \frac{3}{2} \eta
\end{equation}
where $\eta$ is a form factor that describes the shape of a cluster of galaxies through its configuration and the size and mass of each particle's individual halo in comparison to a homogeneous cell. This gives
\begin{equation}\label{eq-Wseta}
W_* = -\frac{2}{3}\frac{\eta}{\zeta(\epsilon/R)}
\end{equation}
where
\begin{equation} \label{eq-etadef}
\eta = \frac{4}{3} \frac{1}{N^2}\sum_{1\leq i < j \leq N} \frac{\kappa(r_{ij})}{r_{ij}/R}.
\end{equation}
This is defined only for cells with $N\geq 2$ because $\eta$ is not meaningful where $N = 0$ or $N = 1$ since there are no other particles in the cell to form a cluster.

\subsubsection{Kinetic Energy}
The kinetic energy of a cell is defined as
\begin{equation}\label{eq-Ktotal}
K = \sum_{i} \frac{1}{2} m v_i^2
 = \frac{1}{2} N m \left\langle v_i^2\right\rangle.
\end{equation}
This depends on $v^2$ which depends on the units used to describe time. A suitable choice for the time unit is a representative dynamical time for the cluster, given by
\begin{equation}\label{eq-tdyn}
\tau_\mathrm{dyn} = (G\rho)^{-1/2} = \sqrt{\frac{R^3}{G m N}}.
\end{equation}
For $v$ given in terms of the dynamical time, the kinetic energy is therefore
\begin{equation}\label{eq-Kdyn}
K = \frac{1}{2} \frac{G N^2 m^2}{R^3}
 \left\langle(v_i\tau_\mathrm{dyn})^2\right\rangle
\end{equation}
and from equation \refeq{eq-Adef} we get
\begin{equation}\label{eq-Tsdef2a}
T_* = \frac{2K}{3NA} = \frac{m\langle v^2 \rangle}{3A}
 = \frac{4}{9}\frac{1}{R^2\zeta(\epsilon/R)} \left\langle(v_i\tau_\mathrm{dyn})^2\right\rangle.
\end{equation}
We can write the velocity as
\begin{equation}\label{eq-Tsups1}
\upsilon^2
 = \frac{1}{R^2}\left\langle(v_i\tau_\mathrm{dyn})^2\right\rangle
\end{equation}
so that $\upsilon^2$ is the mean-squared velocity, given in units of cell radii per dynamical time. In terms of $\upsilon$, we get
\begin{equation}\label{eq-Tsups}
T_* = \frac{4}{9}\frac{\upsilon^2}{\zeta(\epsilon/R)}.
\end{equation}

\subsubsection{Total Energy and Virial Ratio}
\label{sec-Etotal}
The scaled energy of the cell is given by adding equations \refeq{eq-Wseta} and
\refeq{eq-Tsups} to get
\begin{equation}\label{eq-Esdefdim}
E_* = T_* + W_* =
\frac{4}{9}\frac{\upsilon^2}{\zeta(\epsilon/R)} - \frac{2}{3}\frac{\eta}{\zeta(\epsilon/R)}.
\end{equation}
Likewise, the virial ratio $\psi$ is given by the ratio between $W_*$ and $T_*$ such that
\begin{equation}\label{eq-Vrdef2}
\psi = -\frac{W_*}{2T_*} = \frac{3}{4}\frac{\eta}{\upsilon^2}.
\end{equation}

We next compare these quantities to the constraints on $\overline{E}_*$, $\overline{T}_*$ and $\overline{W}_*$ given in section \ref{sec-STP}. The first two constraints are $\overline{T}_* > 0$ and $\overline{W}_* < 0$ which come from the definition of the kinetic and potential energies. The other constraints come from quasi-equilibrium following equations \refeq{eq-EsTs1} and
\refeq{eq-Wsdefa} such that $0 < \overline{\psi} < 1$,
\begin{equation}\label{eq-Wsmin}
\overline{W}_* \geq -\frac{2}{3}2^{2/3} \approx -1.058
\end{equation}
and
\begin{equation}\label{eq-Esmin}
\overline{E}_*
 \geq \frac{\left(\sqrt{10}-4\right)\left(\sqrt{10}-3\right)^{1/3}}{\sqrt{10}-2}
 \approx -0.393.
\end{equation}
The first two constraints are hard limits that also apply to the instantaneous values so that $T_* > 0$ and $W_* < 0$. The next two constraints given in equations \refeq{eq-Wsmin} and \refeq{eq-Esmin} only apply to cells in quasi-equilibrium, but suggest that the values of $W_*$ and $T_*$ are not likely to deviate very far from this range since such cells will quickly relax into a quasi-equilibrium state. This places limits on $\eta$ and $\upsilon$.

The constraint given by the limit on $\overline{W}_*$ in equation \refeq{eq-Wsmin} suggests that for a cell to be in quasi-equilibrium, galaxies cannot be too close to each other because the average pairwise inverse separation $\langle 1/r \rangle$ would become too large. This is closely related to the constraint that $\psi \leq 1$ which comes from the fact that cells where $-W > 2K$ are rapidly collapsing and hence not in quasi-equilibrium or virial equilibrium. Particles in such cells are likely to be close to each other. These particles will experience close encounters resulting in mergers or strong accelerations that substantially change their velocities in a short time. In both cases, the phase space configurations of such cells are highly unstable. This provides a strong constraint on the spatial configuration of a cell, and therefore we expect few cells with arbitrarily dense cores unless they are already relaxed in virial equilibrium.

\subsubsection{Subclusters and mergers}
Galaxies and clusters are known to merge when they get close. Gravitational effects such as dynamical friction, tidal forces and slingshot ejection transfer energy from their orbits into the internal energies of their particles and their orbits merge as their constituent stars or galaxies relax into a single entity.

Early studies of merging galaxies~(e.g. \citealt{1979A&A....76...75R,1980ApJ...236...43A,1996A&A...313..363G}) suggest that the criteria for galaxies to merge depend on the masses and radii of the progenitor galaxies, and their relative velocities, spins, and separations. For example, \citet{1996A&A...313..363G} obtained a merger criterion from empirical fits to $N$-body simulations giving
\begin{equation}\label{eq-mergeGG}
\left[\frac{(m_1+m_2)r_p}{2.5(m_1\epsilon_1+m_2\epsilon_2)}\right]^2 +
\left[\frac{v(r_p)}{1.18v_e(r_p)}\right]^2 \leq 1
\end{equation}
where $m_1$ and $m_2$ are the respective galaxy masses, $\epsilon_1$ and $\epsilon_2$ are their half-mass radii, $r_p$ is the minimum separation between the galaxies, and $v(r_p)$ and $v_e(r_p)$ are their relative velocity and escape velocity at minimum separation. A similar merger criterion by \citet{1980ApJ...236...43A} is
\begin{equation}\label{eq-mergeAF}
\left[\frac{r_p}{2.6(\epsilon_1+\epsilon_2)}\right]^2 +
\left[\frac{v(r_p)}{1.16v_e(r_p)}\right]^2 \leq 1.
\end{equation}
These parameters indicate that galaxy and cluster mergers are complicated events that depend on both the internal and external dynamics of the merging clusters. However, the terms involving $r_p$ in the merger criteria suggest that the separation between galaxies cannot be arbitrarily small or a merger will occur regardless of their mutual velocities. Qualitatively, this means that there is a minimum separation between pairs of galaxies or clusters, and this minimum separation is probably closely related to the masses and radii of both galaxies in the merging pair.

The different merger criteria suggest that there is no simple dependence on the minimum distance between galaxies. Indeed, binary galaxies merge because gravitational effects such as dynamical friction and tidal interactions transfer orbital energy in a tight binary galaxy into the internal kinetic energy of their constituent stars. The efficiency of this process is known to depend on both the dynamics of the encounter and the internal rotation of the merger components~(e.g. \citealt{1972ApJ...178..623T,2010ApJ...725..353D}) and hence a wide variety of conditions may result in a merger. For this reason, we do not quantify the minimum distance between galaxies or subclusters, but note that they are not collisionless and will merge if they get too close to each other. Hence there will be a minimum distance between ``particles'' in a cell.

Even though merging requires time, merging subclusters, being denser than the entire cluster, will have a shorter dynamical time $\tau_\mathrm{dyn}$ than the entire cluster. This means that the merging subcluster can relax to become a single subcluster faster than the overall cluster can relax. Therefore, a stable cluster configuration cannot have more than about 10 subclusters, as we have shown earlier, and these subclusters must be sufficiently spread out that they are unlikely to be merging.

\section{Comparing different shapes}
\label{sec-desc}

From the previous sections, we can narrow the properties of a quasi-equilibrium cluster. Such clusters can be decomposed into subclusters whose masses are within an order of magnitude of each other. Clusters that have a positive specific heat are likely to have less than 10 subclusters, and these subclusters are spread out throughout the cluster. Clusters that have a negative specific heat are more likely to occur than clusters with a positive specific heat, and they are nearly virialized in a mostly closed system. These virialized clusters will have a roughly spherical or ellipsoidal shape, unlike clusters with a positive specific heat which may be more irregular.

The small number of ``particles'' in a cell suggest that an effective description of a cell and hence of a cluster of galaxies is the number and detailed positions of its subclusters. Using this description, compact clusters have a single dominant subcluster, medium compact clusters have multiple subclusters and loose clusters have a single diffuse subcluster that is most likely to be unvirialized.

Based on this classification, compact clusters are best represented as spherical with a density profile. Loose unvirialized clusters are likely to be rare, and have fewer galaxies. Such clusters are better represented as a collection of galaxies. Finally, medium compact clusters, having multiple pronounced concentrations, are best described as a collection of nearly virialized subclusters. With this in mind, we can decompose a rich cluster into a small number of nearly virialized subclusters. These subclusters are simple, nearly spherical objects with density profiles which are easily characterized. Sparser regions of space may be similarly decomposed although they may have different scales.

\subsection{Probability as a function of virial ratio}

From the detailed positions and velocities of the subclusters in a cell, we can calculate the instantaneous energy and virial ratio of the cell. These energies represent a snapshot of a cell that is more closely related to its local dynamics than to its average thermodynamics. The energies fluctuate about the quasi-equilibrium ensemble averages as particles move about the cell. Therefore a measured instantaneous virial ratio corresponds to a range of quasi-equilibrium virial ratios. These fluctuations have been measured to be up to $20\%$ in $N$-body simulations~\citep{1972ApJ...172...17A}, so we use a range of $20\%$ about the measured virial ratio.

This gives us a range over which to integrate equation \refeq{eq-PpsiAbs}, so that the probability that a cell has a measured virial ratio $\psi$ is
\begin{equation} \label{eq-PpsiObs}
P_N(\psi) = \int_{0.8 \psi}^{1.2 \psi} f_V(N)P(\overline{\psi}|N)d \overline{\psi}.
\end{equation}
This probability, although related to the configuration of a cell, is not the probability that a cell has a similar configuration. This is because different configurations may have the same energy and virial ratio. In fact, because galaxies in a cell move about, the detailed spatial configuration of a cell is not static. For this reason, we assume that all configurations with the same virial ratio are equally probable because they are likely to be chance occurrences as galaxies and subclusters move around a cell. Under this assumption, we can compare different configurations and study how much more likely a cell is to have one configuration rather than another.

For a cell in which we can measure both $T_*$ and $W_*$, the measured virial ratio $\psi$ is easily computed. However, for a cell in which we only have the spatial configuration~(e.g. observations with no radial velocity information), we need to make additional assumptions in order to infer a value of $T_*$. Since cells that are not in quasi-equilibrium are likely to relax into quasi-equilibrium rapidly, we may assume that the cell is in quasi-equilibrium to get the most probable value of $T_*$. This means that we can use equations \refeq{eq-Wsdefa} and \refeq{eq-Vrdefa} to estimate $T_*$ and $\psi$ from an observed value of $W_*$. Thus
\begin{equation}\label{eq-WsTsest}
W_* = -\frac{2T_\stx{est.}}{T_\stx{est.}^3+1}
\end{equation}
and
\begin{equation}\label{eq-VrTsest}
\psi_\mathrm{est.} = -\frac{W_*}{2T_\stx{est.}} = \frac{1}{T_\stx{est.}^3+1}
\end{equation}
from which we can use equation \refeq{eq-PpsiObs} to calculate a probability.

Another application of the range of probabilities is to determine if a specific configuration is likely to be bound, unbound or virialized. This is useful for many cases. However, in the case $0.86 \leq \overline{\psi} \leq 0.999$ a cluster has a negative specific heat because most spatial configurations have a value of $\psi$ that falls within the negative specific heat branch of $\overline{E}_*[\overline{\psi}]$ in figure \ref{fig-ivar}. Then the variation in $\psi$ means that we integrate over the entire negative specific heat branch. This suggests that when a cell is virialized, its spatial configuration can take almost any shape.

\subsection{Illustrative cases}

To illustrate this procedure, we compare the two highly idealized configurations of a line and a ring where we focus on the spatial configuration. To reduce the number of independent variables, we ignore the kinetic energy information and use equations \refeq{eq-WsTsest} and \refeq{eq-VrTsest} to estimate the virial ratio. In these configurations, particles are regularly spaced so it is easy to compute the potential energy by a simple summation. The cases are different since the line configuration is a 1-dimensional configuration, and the ring is a flat 2-dimensional configuration.

\subsubsection{Line configuration}
The simplest 1-dimensional configuration is a line with particles spaced at regular intervals. For a cell with $N$ particles and radius $R$, the spacing between each adjacent pair of particles is $2R/(N-1)$. This configuration is infinitesimally thin, and hence the axis ratios are infinite.

Because particles are arranged in a line, particle $i$ has $i-1$ particles to its left and $N-i$ particles to its right. Therefore $W_*$ obtained from equation \refeq{eq-Wseta} is
\begin{equation} \label{eq-Wsline}
W_\stx{line} = -\frac{2}{9} \frac{N-1}{N^2} \sum_{i=1}^{N}\left[\sum_{j=1}^{N-i}\frac{1}{j} + \sum_{j=1}^{i-1}\frac{1}{j} \right].
\end{equation}

\subsubsection{Ring configuration}
A simple case for a 2-dimensional configuration is where particles are arranged in a ring about their common center-of-mass. The particles are spaced evenly around the circumference of a circle so that the distance between particles $i$ and $j$ is $2R \sin(\pi(i-j)/N)$.

From the symmetry of the ring, the total potential energy on any particle in the cluster is the same for all particles. Therefore $W_*$ for the ring configuration is
\begin{equation} \label{eq-Wscirc}
W_\stx{ring} = -\frac{4}{9} \frac{1}{N} \sum_{k=1}^{N-1}\frac{1}{2 \sin(k\pi/N)}.
\end{equation}

\begin{figure}[tbp]
\begin{center}
\includegraphics[width=\floatwidth]{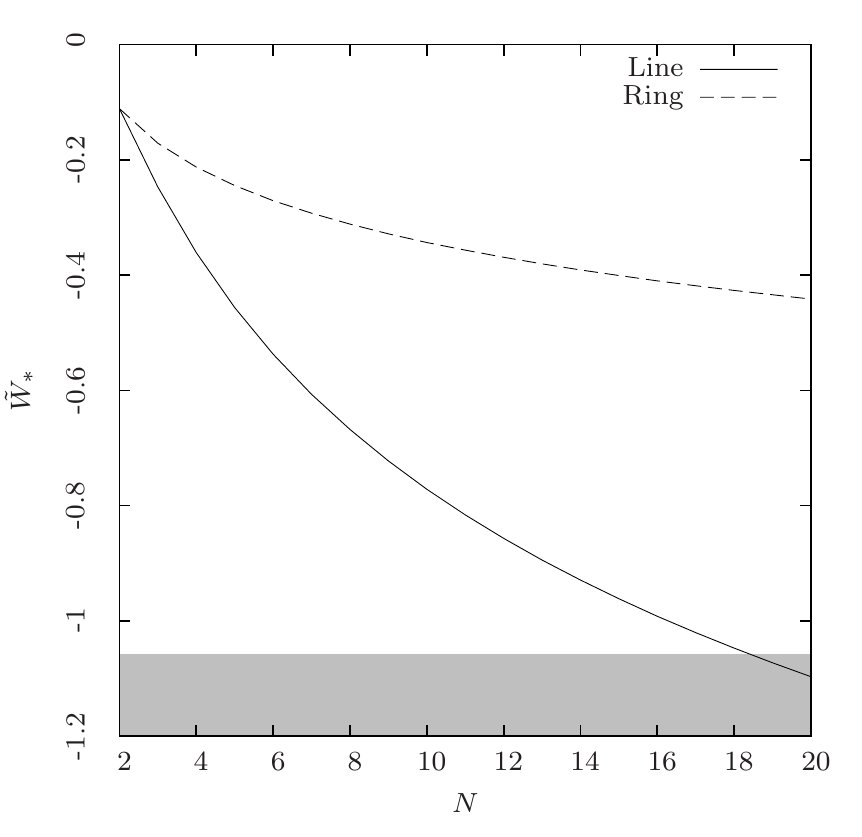}
\caption{
Value of $W_*$ for the line and ring configuration. The shaded area indicates the region where $W_*$ is too negative for the cluster to be in quasi-equilibrium as discussed in section \ref{sec-Etotal}. The value of $W_*$ for both configurations becomes negatively infinite as $N \to \infty$.
}
\label{fig-Wsconf}
\end{center}
\end{figure}

To compare these cases, we plot $W_*$ against $N$ in figure \ref{fig-Wsconf} which shows that that a line configuration cannot be in quasi-equilibrium when $N > 18$. This suggests that filaments cannot be arbitrarily long because filaments that are too long are unstable and would quickly collapse into a different configuration. A similar case can be made for the ring configuration, such that the ring cannot be in quasi-equilibrium when $N > 1563$ but a cell with such a large value of $N$ would have to be very dense or very large.

Very dense cells are unlikely to contain a ring because in a very short dynamical timescale $\tau_{dyn} \sim (G\rho)^{-1/2}$ they would have relaxed into approximate virial equilibrium destroying any quasi-equilibrium ring structure. In addition, a large ring containing $\sim1500$ galaxies is more likely to be primordial than to have been formed through quasi-equilibrium clustering.

Although the ring and line configurations are highly idealized cases, the difference between them indicates that changing the axis ratio of a configuration has an effect on $W_*$. In particular, a 2-dimensional configuration allows for more distance between particles than a 1-dimensional case, and hence a larger nearest-neighbor distance. This makes $W_*$ less negative for the 2-dimensional case than the 1-dimensional cases since $W_*$ depends on the average inverse separation between particles.

This dependence on dimensionality illustrates how shape influences the potential energy of a cluster. For example, in the simplest case of a uniform cluster, a spheroidal cluster is likely to have a more negative value of $W_*$ than a spherical cluster of the same mass and major axis. However, there are many shapes that have the same energy although it is possible to calculate the potential energy from the shape and distribution of galaxies in a cluster. Thus the internal structure of a cluster plays an important role in characterizing the potential energy of a cluster and a simple description of shape is likely to be insufficient to describe a cluster. Therefore we use the detailed positions of the galaxies or subclusters instead.

\begin{figure*}[tbp]
\begin{center}
\includegraphics[width=\floatwidth]{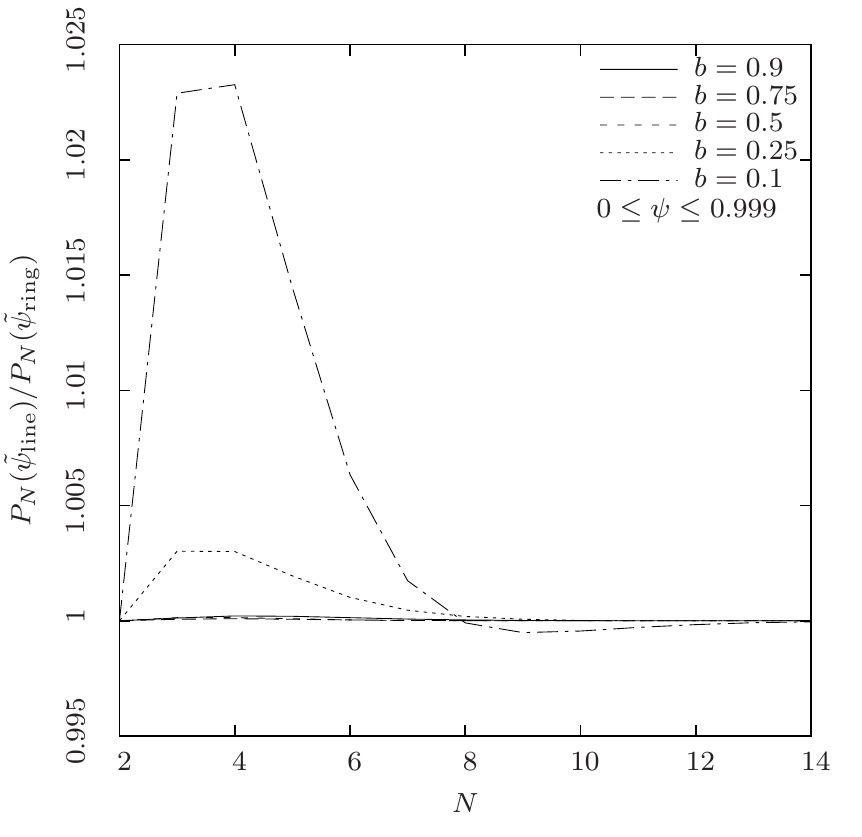}
\includegraphics[width=\floatwidth]{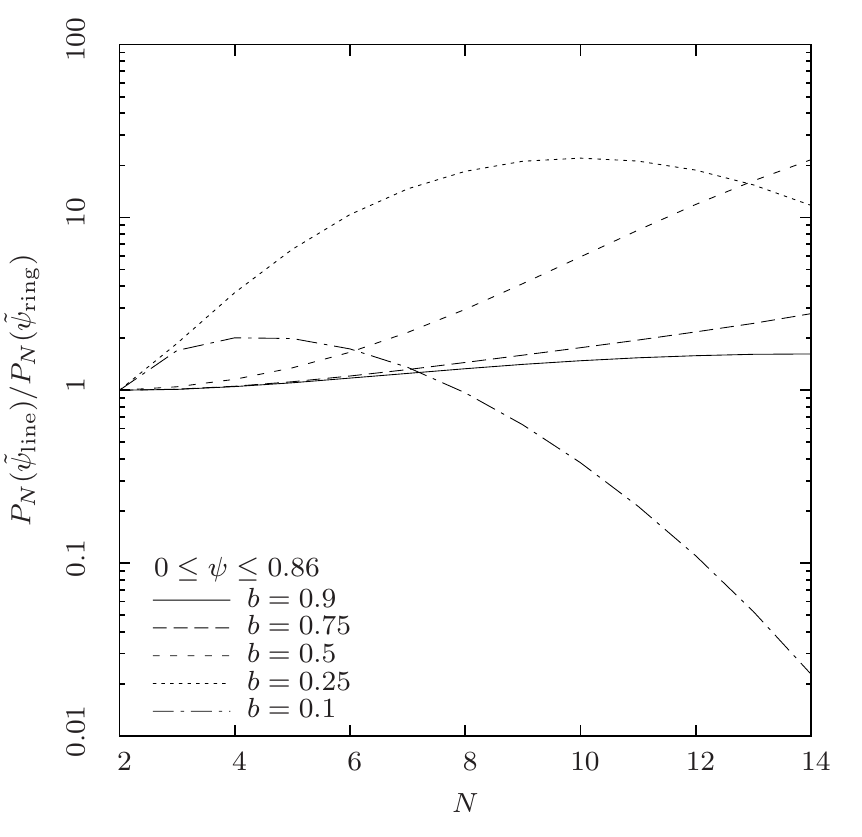}\\
\includegraphics[width=\floatwidth]{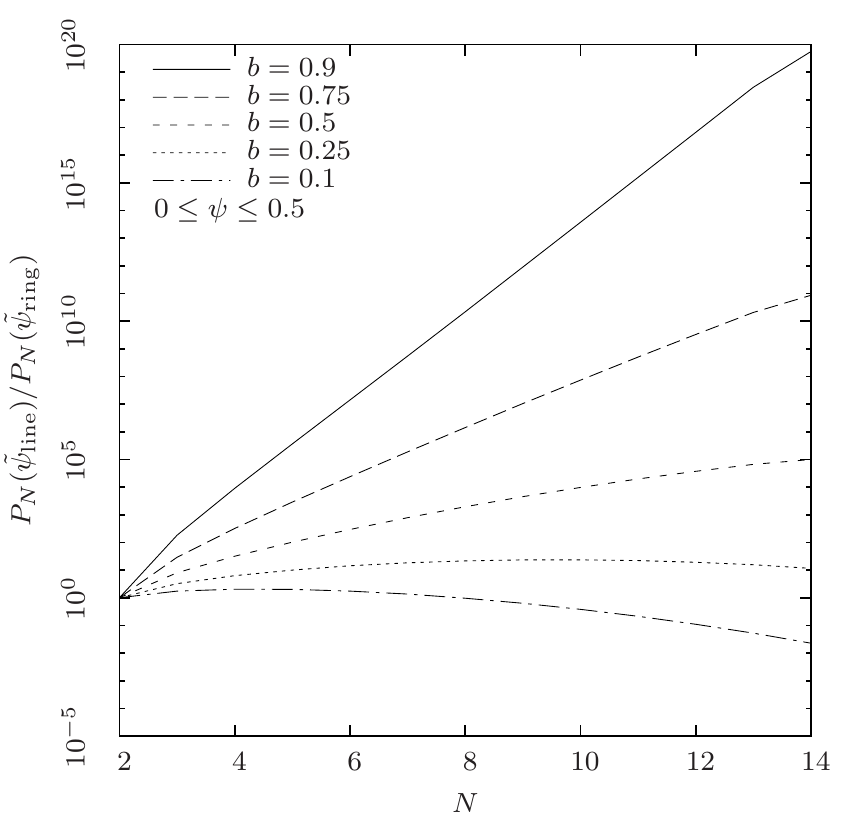}
\includegraphics[width=\floatwidth]{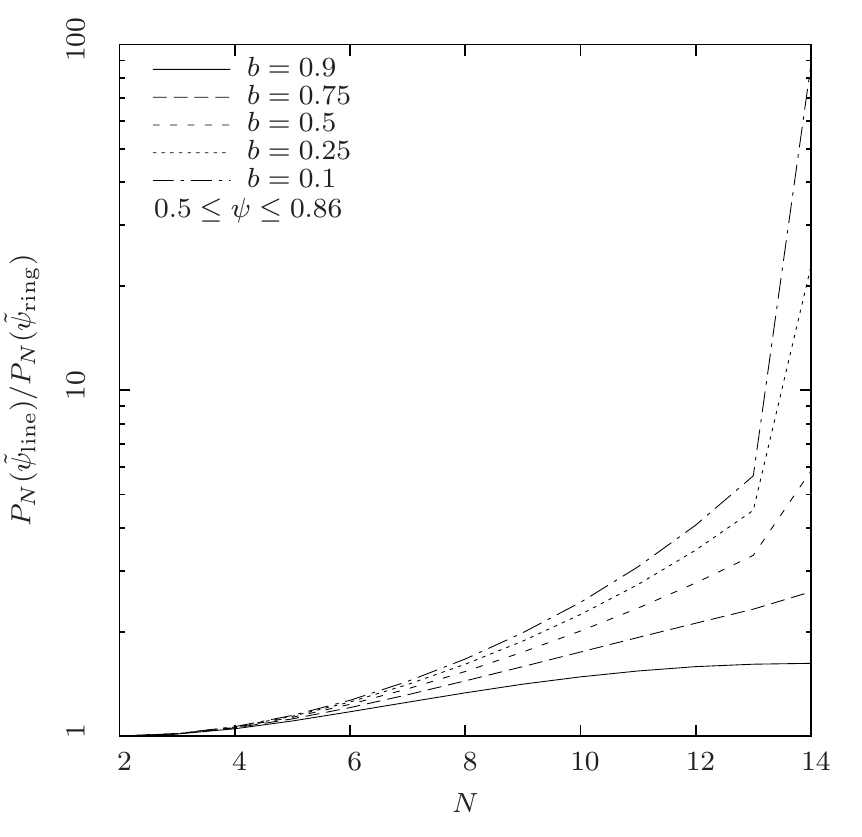}
\caption{
Relative probability that a cell has a line configuration rather than a ring configuration for different conditions.
Top left: Cells in quasi-equilibrium. Top right: Cells with a positive specific heat. Bottom left: Unbound cells. Bottom right: Bound cells with a positive specific heat.
}
\label{fig-PShape}
\end{center}
\end{figure*}

To compare the line and the ring configuration, we calculate an estimate of $\psi$ for both configurations using equations \refeq{eq-WsTsest} and \refeq{eq-VrTsest} where $W_*$ is given by equations \refeq{eq-Wsline} and \refeq{eq-Wscirc} for the line and ring respectively. Using these values of $\psi$, we use equation \refeq{eq-PpsiObs} to determine the probability that a cell has a similar virial ratio which is a necessary condition to have a specific configuration. To compare these two shapes, we take the ratio of the probabilities $P_N(\psi_\mathrm{line})/P_N(\psi_\mathrm{ring})$ which tells us how much more likely a cell is to contain a line than a ring. This ratio is independent of $\overline{N}$ because the $f_V(N)$ term cancels out in the ratio. We plot $P_N(\psi_\mathrm{line})/P_N(\psi_\mathrm{ring})$ for different regimes of $\psi$ in figure \ref{fig-PShape}.

Here, we do not plot the comparison for the negative specific heat branch because it covers a very small range of $\overline{\psi}$ such that any configuration with a value of $\psi$ in the negative specific heat branch will include the entire negative specific heat branch because of the roughly $20\%$ statistical variations over time of $\psi$ in equation \refeq{eq-PpsiObs}. In addition, $W_*[\psi]$ is double valued with a minimum at $\psi = 2/3$, so for almost every value of $W_*$ there is a value of $\psi$ in both the positive specific heat and negative specific heat branches. This means that almost any spatial configuration is equally likely in the negative specific heat branch because there always exists a value of $\psi$ such that the integral in equation \refeq{eq-PpsiObs} covers the entire negative specific heat branch. The only spatial configurations that are unlikely to occur in the negative specific heat branch are those with very negative values of $W_*$. Since the absolute value of $W_*$ depends on the inverse separation between particles, such configurations will have particles that are very close to each other, and therefore are likely to be unstable to mergers.

From figure \ref{fig-PShape}, in many cases the line configuration is more probable than the ring configuration, and it is only in the case of small $b$ or large $N$ that a ring configuration is more probable than a line configuration. This means that in most of the universe, we are more likely to find galaxies arranged in a line than galaxies arranged in a ring. This prediction is consistent with the observation that filaments are more frequently observed than galaxies or clusters arranged in a ring.

\section{Conclusions and Discussion}
\label{sec-conc}

In this paper we have described a theory of clustering that extends the work of \citet{2004ApJ...608..636L} to calculate the probability that a cell has a given time-averaged virial ratio $\overline{\psi}$. This probability is independent of the average number of galaxies in a cell and depends on the clustering parameter $b$, the number of galaxies $N$ in the cell and the range of virial ratios that the cell may have.

A feature of this theory is the use of subclusters to describe the internal spatial configuration of a cell. These subclusters are concentrations of galaxies that are nearly virialized and share a common bulk velocity. The use of subclusters considerably simplifies the analysis because their masses are approximately equal and most clusters that are still assembling will have less than 10 subclusters. This is because cells with more than 10 subclusters are likely to already be virialized, and subclusters that are less massive than a tenth of the most massive subcluster do not have enough mass to significantly affect the potential energy.

Because subclusters are collections of galaxies, the number of subclusters within a single cluster can be an indicator of its evolutionary stage. In hierarchical models, clusters build up by the merger of multiple subclusters. Thus the presence of multiple clusters within a cell, or significant substructure within a cluster would suggest that mergers were relatively recent. Such clusters, having multiple subclusters, are classified as medium compact clusters in the classification scheme of \citet{1957PASP...69..409H}. More evolved relaxed clusters have less substructure with a single dominant concentration. Such clusters are their own subcluster and are classified as compact clusters under the same classification scheme.

In addition, clusters with a negative specific heat may take on almost any configuration because $W_*[\psi]$ is double-valued with a minimum at $\psi=2/3$. Hence for almost every value of $W_*$ there is a value of $\psi$ in the negative specific heat branch. The only spatial configurations that are unlikely to occur in the negative specific heat branch are those with very negative values of $W_*$. Since the absolute value of $W_*$ depends on the inverse separation between particles, such configurations will have particles that are very close to each other, and therefore are likely to be unstable to mergers. Together with the roughly $20\%$ statistical variation over time of $\psi$ about $\overline{\psi}$, any configuration with a value of $\psi$ in the negative specific heat branch may be a chance configuration of a cluster that has a negative specific heat.

Since the $6N$-dimensional phase space configuration of subclusters in a cell is directly related to the cell's potential and kinetic energies and thus its virial ratio, we can use the theory to study the internal structure and shapes within a cell. To do this, we relate the energy calculated from the observed instantaneous phase space configuration to the ensemble average energies in quasi-equilibrium. This highlights the fact that clusters of galaxies are not in strict equilibrium and hence their potential and kinetic energies fluctuate about the ensemble average energy. The spatial configurations of clusters change as a result of these fluctuations.

Because the average thermodynamic quantities of an ensemble of cells in quasi-equilibrium change very slowly compared to the dynamical timescale of a single cell, the quasi-equilibrium energies are approximately time averages. We use the spectrum of fluctuations to determine the probability that a cell may have a particular instantaneous virial ratio. This is a necessary but not sufficient condition for a cell to have a given configuration. To compare different configurations, we therefore compare the probability that a cell has a virial ratio which could give rise to a given configuration.

Based on the range of fluctuations in virial ratio observed in $N$-body simulations by \citet{1972ApJ...172...17A}, we conclude that cells that have a negative specific heat may have almost any spatial configuration. These configurations are likely to be chance configurations of fluctuating cells that are nearly virialized. Cells with positive specific heat have a range of energies corresponding to fewer spatial configurations. In our comparisons between line and ring configurations, the probability that a cell will have a line configuration rather than a ring configuration varies depending on the clustering parameter $b$ and the number $N$ of subclusters in the cell. Therefore we present the following procedure for comparing the shapes of two clusters:

The first step is to define a region of space that covers the cluster. The size and shape of this region of space define the cell size. Within the cell, identify the subclusters whose masses are within an order of magnitude of the largest subcluster. The positions of these subclusters are directly related to the scaled potential energy $W_*$ of the cluster. From the observed $W_*$ or the peculiar velocity information, estimate the virial ratio $\psi$.

The next step is to use the cell size to calculate the mean and variance of the counts-in-cells distribution of similarly-sized subclusters for a larger sample to obtain the clustering parameter $b$~(e.g. \citealt{2005ApJ...626..795S,2011ApJ...729..123Y}). Then using the observed parameters $\psi$ and $b$, calculate the probability of finding a cell with a similar virial ratio using equation \refeq{eq-PpsiObs}. This probability should then be compared to a reference cell that has a different structure to determine the relative probabilities of different configurations.

Although the configurations that we use here are highly idealized, our comparison of a line and a ring configuration is generally consistent with observations of the cosmic web. This simple test of our theory gives a reasonable result. We intend to make more detailed comparisons with observations and more realistic cases which may include dark matter in a forthcoming paper.

\acknowledgements
We wish to acknowledge the preliminary work of Chuah Boon Leng who helped explore some of the concepts discussed in this paper. We also wish to thank Phil Chan and Bernard Leong for many helpful discussions on this topic.

\appendix
\section{Multiple mass components}
\label{app-mmass}
To illustrate the effect of having multiple mass components in a cluster, we consider the case where a cluster has two populations of subclusters. For simplicity, we assume that these subclusters are point masses, and the masses of each subcluster are $m_1$ and $m_2$ with $m_2 > m_1$ for subclusters in each population.

In such a cluster, we can consider the interactions of particles from each population such that the total potential of the cluster is~(c.f. Equation \ref{eq-PEBinary})
\begin{equation}\label{eq-PE2comp}
W = -G\left[
   \frac{1}{2}\sum_{i\neq j}\frac{m_1^2} {|\mathbf{x}_i^{(1)}-\mathbf{x}_j^{(1)}|}
  +\frac{1}{2}\sum_{i\neq j}\frac{m_2^2} {|\mathbf{x}_i^{(2)}-\mathbf{x}_j^{(2)}|}
  + \sum_{i^{(1)}, j^{(2)}}\frac{m_1 m_2} {|\mathbf{x}_i^{(1)}-\mathbf{x}_j^{(2)}|}\right]
\end{equation}
where the superscripts denote members of the different populations, and $\mathbf{x}$ denotes position. The three terms in equation \refeq{eq-PE2comp} come from considering each mass component as a separate system, and adding their mutual interactions. Therefore, equation \refeq{eq-PE2comp} is similar in form to equation \refeq{eq-PEBinary}, even though we do not place any constraints on the positions of individual particles.

We can take the average of the sums in equation \refeq{eq-PE2comp} such that
\begin{equation}\label{eq-PE2cavg}
W = -G\left[
  m_1^2\frac{N_1(N_1-1)}{2}\left\langle \frac{1}{r^{(1)}}\right\rangle 
  + m_2^2\frac{N_2(N_2-1)}{2}\left\langle \frac{1}{r^{(2)}}\right\rangle
  + m_1 m_2 (N_1 N_2)\left\langle \frac{1}{r^{(1,2)}}\right\rangle
 \right]
\end{equation}
where $N_1$ and $N_2$ are the numbers of particles in each population, and $\langle 1/r^{(1)} \rangle$ is the average inverse separation between members of population 1. In the case where both populations are similarly distributed throughout the cell, the separate averages are approximately equal, i.e. $\langle 1/r \rangle \approx \langle 1/r^{(1)} \rangle \approx \langle 1/r^{(2)} \rangle \approx \langle 1/r^{(1,2)} \rangle$ so only the mass and number ratios affect the total potential.

Therefore, equation \refeq{eq-Wsdef2} gives~(c.f. \citealt{2006IJMPD..15.1267A})
\begin{equation}\label{eq-Ws2cavg}
W_\stx{2comp.} = - \frac{4}{9} \left\langle\frac{1}{r}\right\rangle
  R \left[ 1- \frac{N_1+N_2 \left(m_2/m_1\right)^2}{\left[N_1+N_2 \left(m_2/m_1\right)\right]^2} \right]
\end{equation}
which reduces to the single mass case given by equation \refeq{eq-Wsdef2} for $m_1 = m_2$ or $N_2 = 0$. When the mass ratio $m_2/m_1$ is sufficiently large that the $N_2(m_2/m_1)$ term is larger than the $N_1$ term in the denominator, we can ignore the less massive component. In such cases, most of the total mass is in the more massive particles and these particles dominate the total potential energy of the cell. 

This means that in cases where most of the mass in a cell is concentrated in a single large subcluster, the large subcluster dominates the potential and all the other smaller subclusters are satellites around the large subcluster. In such cells, the theory we have described thus far does not apply because there are no other similar subclusters for the dominant subcluster to cluster with.

\begin{figure*}[tbp]
\begin{center}
\includegraphics[width=\floatwidth]{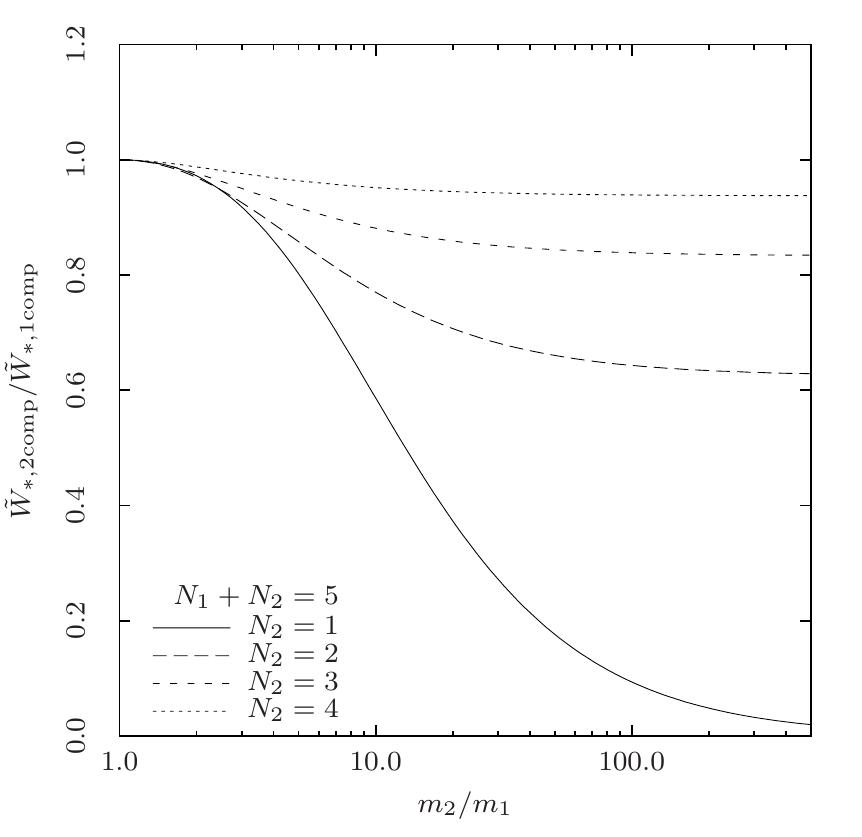}
\includegraphics[width=\floatwidth]{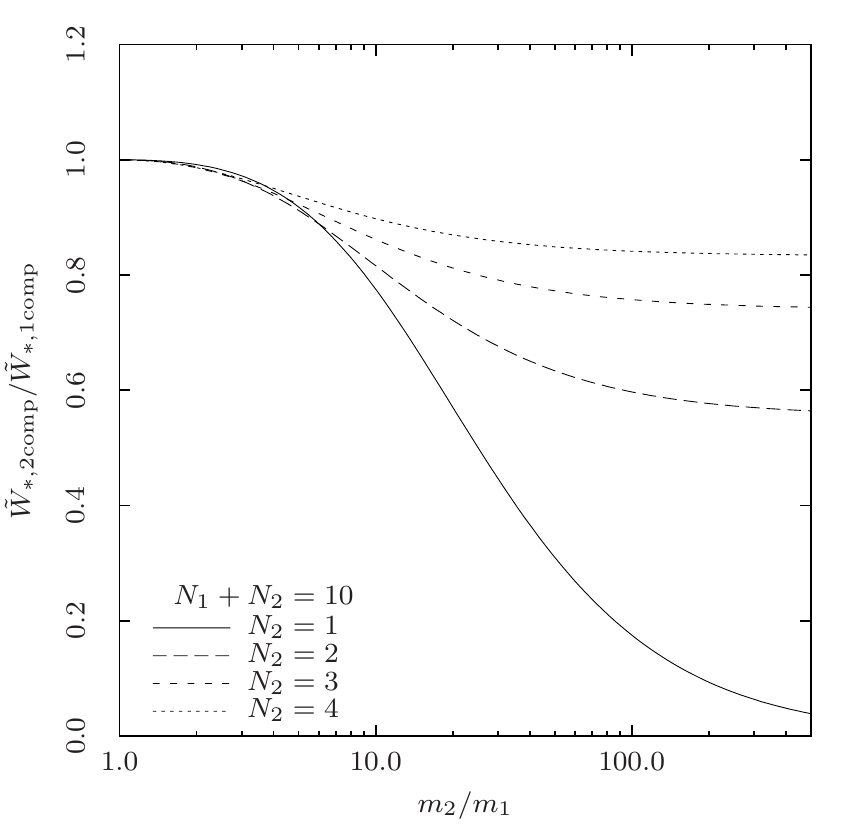}
\caption{
$W_{*,2comp}/W_{*,1comp}$ for different total numbers of particles $N$, numbers of more massive particles $N_2$ and mass ratio $m_2/m_1$. Left: $N = 5$, right: $N = 10$.
}
\label{fig-2compWs}
\end{center}
\end{figure*}

For a cluster with less than 10 subclusters, a mass ratio of $(m_2/m_1) \approx 10$ is large enough such that $N_2(m_2/m_1) > N_1$. This means that we only need to consider subclusters that are at least a tenth as massive as the most massive subcluster in the cluster. These subclusters will contain most of the mass in the cluster, and less massive subclusters are likely to be satellites around these more massive subclusters. To illustrate this, we compare $W_\stx{2comp.}$ with the single mass case for different mass ratios and $N_2$ in figure \ref{fig-2compWs}. In the two population case, $W_*$ is lower than in the single mass case. At $m_2/m_1 \gtrsim 10$, the difference may be larger than $20\%$ depending on the number ratios. This means that particles that make a significant contribution to the total potential energy will have the same mass to within an order of magnitude.

\begin{thebibliography}{24}
\expandafter\ifx\csname natexlab\endcsname\relax\def\natexlab#1{#1}\fi

\bibitem[{{Aarseth} \& {Fall}(1980)}]{1980ApJ...236...43A}
{Aarseth}, S.~J. and {Fall}, S.~M. 1980, \apj, 236, 43

\bibitem[{{Aarseth} \& {Saslaw}(1972)}]{1972ApJ...172...17A}
{Aarseth}, S.~J. and {Saslaw}, W.~C. 1972, \apj, 172, 17

\bibitem[{{Abell}(1958)}]{1958ApJS....3..211A}
{Abell}, G.~O. 1958, \apjs, 3, 211

\bibitem[{{Ahmad} {et~al.}(2006{\natexlab{a}}){Ahmad}, {Malik}, \&
  {Masood}}]{2006IJMPD..15.1267A}
{Ahmad}, F., {Malik}, M.~A., and {Masood}, S. 2006{\natexlab{a}}, International
  Journal of Modern Physics D, 15, 1267

\bibitem[{{Ahmad} {et~al.}(2002){Ahmad}, {Saslaw}, \&
  {Bhat}}]{2002ApJ...571..576A}
{Ahmad}, F., {Saslaw}, W.~C., and {Bhat}, N.~I. 2002, \apj, 571, 576

\bibitem[{{Ahmad} {et~al.}(2006{\natexlab{b}}){Ahmad}, {Saslaw}, \&
  {Malik}}]{2006ApJ...645..940A}
{Ahmad}, F., {Saslaw}, W.~C., and {Malik}, M.~A. 2006{\natexlab{b}}, \apj, 645,
  940

\bibitem[{{Baumann} {et~al.}(2003){Baumann}, {Leong}, \&
  {Saslaw}}]{2003MNRAS.345..552B}
{Baumann}, D., {Leong}, B., and {Saslaw}, W.~C. 2003, \mnras, 345, 552

\bibitem[{{Binggeli} {et~al.}(1987){Binggeli}, {Tammann}, \&
  {Sandage}}]{1987AJ.....94..251B}
{Binggeli}, B., {Tammann}, G.~A., and {Sandage}, A. 1987, \aj, 94, 251

\bibitem[{{D'Onghia} {et~al.}(2010){D'Onghia}, {Vogelsberger},
  {Faucher-Giguere}, \& {Hernquist}}]{2010ApJ...725..353D}
{D'Onghia}, E., {Vogelsberger}, M., {Faucher-Giguere}, C., and {Hernquist}, L.
  2010, \apj, 725, 353

\bibitem[{{Garcia-Gomez} {et~al.}(1996){Garcia-Gomez}, {Athanassoula}, \&
  {Garijo}}]{1996A&A...313..363G}
{Garcia-Gomez}, C., {Athanassoula}, E., and {Garijo}, A. 1996, \aap, 313, 363

\bibitem[{{Herzog} {et~al.}(1957){Herzog}, {Wild}, \&
  {Zwicky}}]{1957PASP...69..409H}
{Herzog}, E., {Wild}, P., and {Zwicky}, F. 1957, \pasp, 69, 409

\bibitem[{{Itoh} {et~al.}(1988){Itoh}, {Inagaki}, \&
  {Saslaw}}]{1988ApJ...331...45I}
{Itoh}, M., {Inagaki}, S., and {Saslaw}, W.~C. 1988, \apj, 331, 45

\bibitem[{{Landau} \& {Lifshitz}(1980)}]{1980stph.book.....L}
{Landau}, L.~D. and {Lifshitz}, E.~M. 1980, {Statistical Physics} (Pergamon
  Press, Oxford)

\bibitem[{{Leong} \& {Saslaw}(2004)}]{2004ApJ...608..636L}
{Leong}, B. and {Saslaw}, W.~C. 2004, \apj, 608, 636

\bibitem[{{Roos} \& {Norman}(1979)}]{1979A&A....76...75R}
{Roos}, N. and {Norman}, C.~A. 1979, \aap, 76, 75

\bibitem[{{Saslaw} \& {Ahmad}(2010)}]{2010ApJ...720.1246S}
{Saslaw}, W.~C. and {Ahmad}, F. 2010, \apj, 720, 1246

\bibitem[{{Saslaw} {et~al.}(1990){Saslaw}, {Chitre}, {Itoh}, \&
  {Inagaki}}]{1990ApJ...365..419S}
{Saslaw}, W.~C., {Chitre}, S.~M., {Itoh}, M., and {Inagaki}, S. 1990, \apj,
  365, 419

\bibitem[{{Saslaw} \& {Fang}(1996)}]{1996ApJ...460...16S}
{Saslaw}, W.~C. and {Fang}, F. 1996, \apj, 460, 16

\bibitem[{{Saslaw} \& {Hamilton}(1984)}]{1984ApJ...276...13S}
{Saslaw}, W.~C. and {Hamilton}, A.~J.~S. 1984, \apj, 276, 13

\bibitem[{{Saslaw} \& {Yang}(2010)}]{2009arXiv0902.0747S}
{Saslaw}, W.~C. and {Yang}, A. 2010, in Lecture Notes of the Les Houches Summer
  School: Long-Range Interacting Systems, ed. T.~{Dauxois}, S.~{Ruffo}, \&
  L.~F. {Cugliandolo}, Vol.~XC (Oxford University Press, Oxford, UK), 377--398

\bibitem[{{Sivakoff} \& {Saslaw}(2005)}]{2005ApJ...626..795S}
{Sivakoff}, G.~R. and {Saslaw}, W.~C. 2005, \apj, 626, 795

\bibitem[{{Toomre} \& {Toomre}(1972)}]{1972ApJ...178..623T}
{Toomre}, A. and {Toomre}, J. 1972, \apj, 178, 623

\bibitem[{{Yang} \& {Saslaw}(2011)}]{2011ApJ...729..123Y}
{Yang}, A. and {Saslaw}, W.~C. 2011, \apj, 729, 123

\bibitem[{{Yang} {et~al.}(2011){Yang}, {Saslaw}, {Chan}, \&
  {Leong}}]{2010arXiv1011.0176Y}
{Yang}, A., {Saslaw}, W.~C., {Chan}, A.~H., and {Leong}, B. 2011, in
  Proceedings of the Conference in Honour of Murray Gell-Mann's 80th Birthday,
  ed. H.~Fritzsch \& K.~K. Phua (World Scientific, Singapore), 597--604

\end{thebibliography}
\end{document}